\documentclass[11pt]{article}

\usepackage[final]{acl}

\usepackage{times}
\usepackage{latexsym}

\usepackage[T2A,T1]{fontenc}
\usepackage[T5]{fontenc}

\usepackage[utf8]{inputenc}

\usepackage{microtype}

\usepackage{inconsolata}

\usepackage{graphicx}

\usepackage[most]{tcolorbox}
\usepackage{enumitem}
\newcommand{\cyrtext}[1]{{\fontencoding{T2A}\selectfont #1}}
\usepackage{float}
\usepackage{booktabs}
\usepackage{tabularx}
\usepackage{xcolor}

\newtcolorbox{qualexample}[2][]{
    colback=gray!5,
    colframe=black!60,
    fonttitle=\bfseries,
    title=#2,
    breakable,
    boxrule=0.5pt,
    left=4pt, right=4pt, top=3pt, bottom=3pt,
    #1
}

\newcommand{\gain}[1]{%
  \ifdim #1pt>0pt \textcolor{green!60!black}{+#1}%
  \else\ifdim #1pt<0pt \textcolor{red}{#1}%
  \else #1%
  \fi\fi}

\usepackage{listings}
\tcbuselibrary{listings}
\usepackage{seqsplit}
\usepackage{xurl}

\newcolumntype{Y}{>{\centering\arraybackslash}X}

\title{An Actionable Diagnosis of Multilingual, Multi-Agent Planning Failures}

\author{
  \textbf{Vikas Pahuja\textsuperscript{*}},
  \textbf{Jonathan Brokman\textsuperscript{*}},
  \textbf{Omer Hofman\textsuperscript{1}},
  \textbf{Tamir Nizri\textsuperscript{1}},
  \textbf{Daniel Vishna\textsuperscript{1}},
  \\
  \textbf{Seraphina Goldfarb-Tarrant\textsuperscript{2}},
  \textbf{Kelly Marchisio\textsuperscript{2}},
  \textbf{Hisashi Kojima\textsuperscript{3}},
  \textbf{Roman Vainshtein\textsuperscript{1}}
  \\
  \\
  \textsuperscript{1}Fujitsu Research of Europe
  \qquad
  \textsuperscript{2}Cohere
  \qquad
  \textsuperscript{3}Fujitsu Research
  \\
  \textsuperscript{*}Equal contribution.
}

\begin{document}
\maketitle
\begin{abstract}

Multilingual multi-agent systems exhibit substantial degradation beyond English, yet prior work rarely identifies how task-critical information is lost when user requests are converted into executable plans. We study the planner in a multi-agent system as the request-to-action interface and derive an actionable taxonomy of planning-grounding failures from failed real-world task executions. LLM-based analysis shows that these failures constitute an increasing share of unsuccessful executions as language-resource availability declines, with the strongest effects in low-resource languages. To test whether the taxonomy supports mitigation, we introduce TART, Taxonomy-Guided Actionable Representation, that makes the taxonomy's key aspects explicit to the planner and downstream sub-agents. Across multiple languages, three LLM backbones, two datasets, and two agentic configurations, TART consistently improves performance. On multilingual GAIA, it raises a state-of-the-art system's accuracy by 5.6 percentage points averaged across eleven languages spanning low- to high-resource settings. 
\end{abstract}

\section{Introduction}

\begin{figure*}[t]
  \includegraphics[width=2\columnwidth]{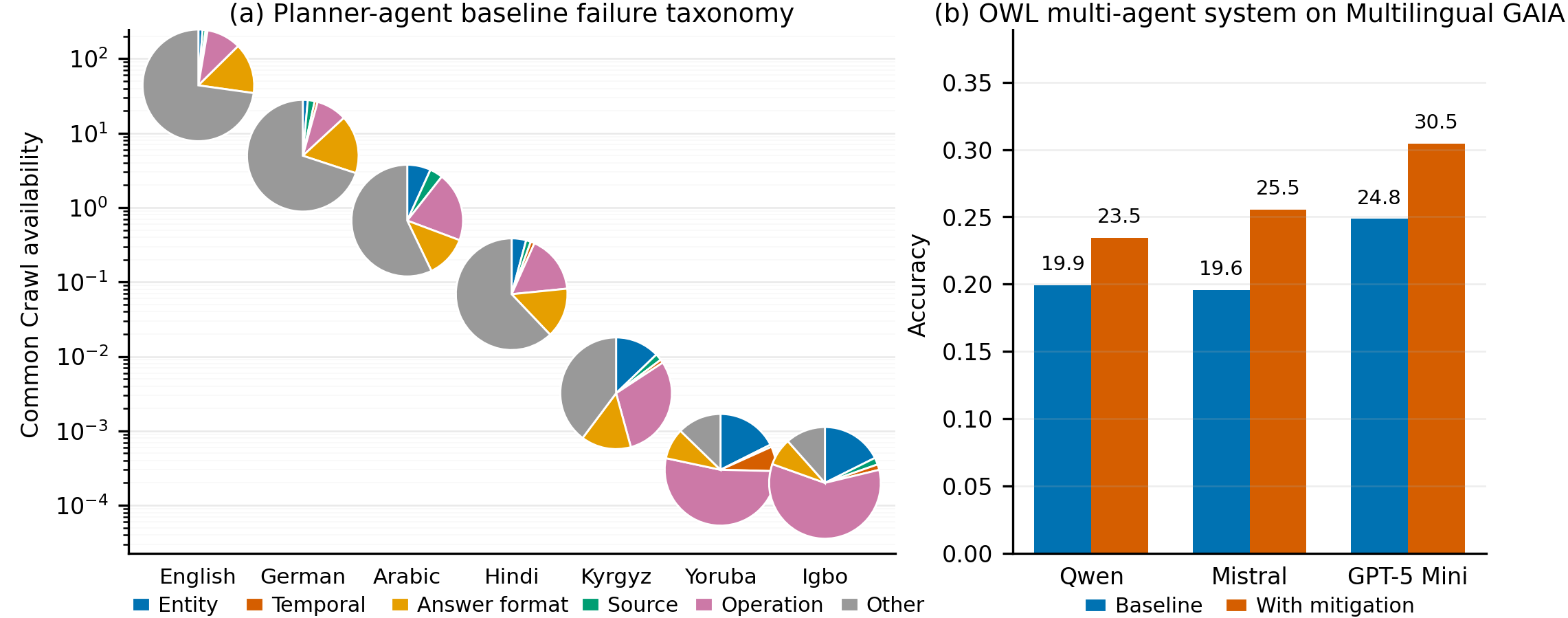}
  \caption{We propose a taxonomy for multi-agent planning failures, developed to diagnose multilingual settings, and guide actionable improvements (taxonomy summarized in Fig.~\ref{fig:taxonomy}). (a) Distributions of the taxonomy's planning failures, measured per-language and ordered by Common Crawl language availability (high to low resource). As language-resource availability decreases, taxonomy-covered failures account for an increasing share of unsuccessful executions—particularly operation and entity grounding—while answer formatting failures decrease. (b) Accuracy of OWL on Multilingual GAIA, averaged over the languages. TART (our method) improves all three configurations. Together, the panels connect a systematic multilingual failure pattern to an actionable mitigation that improves a state-of-the-art multi-agent system.}
  \label{fig:experiments}
\end{figure*}

\begin{figure*}[t]
  \centering
  \includegraphics[width=1.0\textwidth]{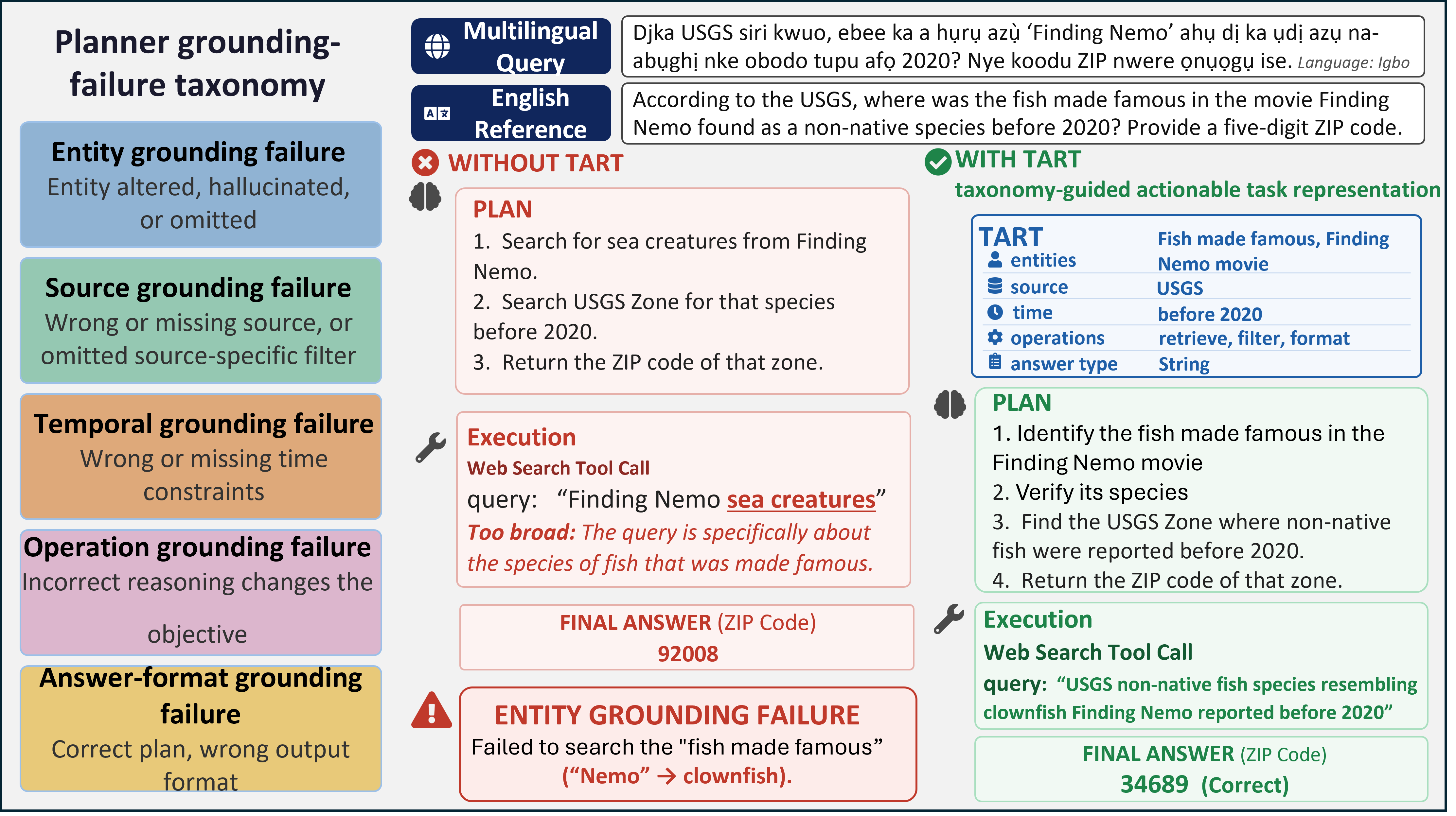}
  \caption{
  \textbf{Left:} Planner-specific multilingual grounding-failure taxonomy. \textbf{Right:} Entity grounding failure example for an Igbo query. Semantic task representation makes the implicit entity explicit and guides both planning and tool use toward the correct USGS retrieval target.
  }
  \label{fig:entity-grounding-example} \label{fig:taxonomy}
\end{figure*}


Language-model-based multi-agent systems are increasingly used as general-purpose tools by a linguistically diverse user base. They translate multilingual user requests into actionable plans, coordinate specialized agents, invoke tools, and generate answers. Their reliability depends on preserving user goals and constraints across languages with unequal representation in training data, making cross-language reliability a significant challenge \cite{joshi2020state, wenzek2020ccnet}.


Recent work has consistently documented cross-lingual degradation in LLM-based agents, from structured function-call generation \cite{kulkarni2025massive,luo2026lost}, through controlled web interaction \cite{wang2025x}, to complex tasks derived from GAIA in multilingual settings \citep[GAIA-MAPS;][]{hofman2026maps,kim2026gaia}. 
However, these studies do not systematically characterize internally which task-critical commitments are lost or distorted inside the multi-agent system.

The request-to-plan boundary is of particular importance: Firstly - it is often the de-facto interface to multilingual inputs, even when following actions are done internally in English or other representations. Moreover, a misunderstanding here is particularly consequential since any distortion might propagate through the entire multi-agent workflow. We define a \emph{planning-grounding failure} as a mismatch between the user request and the generated plan such that, even if the plan were executed correctly, the resulting execution would not satisfy the original request. For example, a plan may substitute the target entity, omit a required source or date
, or follow the wrong sequence of operations. 


Through qualitative analysis
, we identify five recurring forms of planning-grounding failure: entity, source, temporal, operation, and output grounding. These categories correspond to five semantic commitments that a plan must preserve. Moreover, they are directly actionable: identifying these does not merely mean we should re-plan, but guides \textit{how} to re-plan.
After identification, the failure categories were verified to recur across LLMs, agents and languages. Furthermore, LLM-based quantification reveals a broad resource-related gradient: they account for an increasingly large share of unsuccessful executions as language-resource availability decreases, with the pattern already visible in medium-resource settings and most pronounced in lower-resource languages.

We then ask whether the taxonomy is actionable rather than merely descriptive. Guided by its five dimensions, we introduce a TART: Taxonomy-guided Actionable Task Representation. It is a semantic parsing protocol that represents outputs plan-grounding information before planning begins. This structured representation is designed to reflect the proposed taxonomy of errors. It acts as a semantic contract between the user request and the agentic pipeline
, helping the system preserve the commitments of the request. The protocol requires neither changes to the underlying model parameters nor a replacement of the multi-agent framework.

We evaluate TART primarily on multilingual versions of GAIA \cite{mialon2024gaia, hofman2026maps, kim2026gaia}, containing complex, real-world agentic tasks. We extend this setting with additional lower-resource languages, yielding a spectrum from English and higher-resource languages to medium- and lower-resource settings. We test on and additional dataset, MULTITAT as a more controlled setting \cite{zhang2025multitat}. 

On GAIA-MAPS, TART improves a SOTA system, OWL \cite{owl2025}, with GPT-5-mini by 5.6 percentage points, averaged across eleven languages. We further observe improvements with Mistral-Large-3 and Qwen3-VL-235B-A22B
.



To our knowledge, this is the first work to characterize planning failures across languages, and the first  to derive and evaluate a mitigation directly from such a taxonomy.\footnote{Official implementation: \url{https://anonymous.4open.science/r/Multilingual-Multi-Agent-TART-0B42/}.} Main contributions:
\begin{itemize}
    \item An actionable taxonomy of planning-grounding failures that captures which semantic commitments that are lost or distorted when multilingual user requests are converted into plans. 
    Quantification across languages identifies a broad language-resource-related gradient.

    \item Demonstration of the taxonomy's actionability by proposing TART - a protocol that explicitly reflects our taxonomy in guiding planning and downstream coordination. Empirical evidence across languages, LLM backbones and datasets shows that TART improves the end-to-end performance of multi-agent systems on complex multilingual GAIA-style tasks
    .
\end{itemize}

\section{Related Work}
While LLM literature is mature, only recently have studies shown the language effects on LLM-based agent performance. X-WebAgentBench \cite{wang2025x} is a simplified version of an e-commerce web shop
. They evaluate multilingual agents and multi-agent systems, and test 
 alignment methods. 
Single-turn structured function calls have been studied in MASSIVE, providing large-scale evidence that these are sensitive to language 
 \cite{kulkarni2025massive}. \textcolor{black}{MLCL \cite{luo2026lost} similarly studies single function LLM calls
, concerning single structured tool calls
.} 
MULTITAT \cite{zhang2025multitat} focuses on multilingual table reasoning, by translating several table reasoning tasks \cite{chen2020hybridqa, zhu2021tat,zhang2025scitat} into 11 languages, it requires agents to read tables and perform several-step operations on them, making it a simple yet highly relevant use case for multi-step plan evaluation. 
\textcolor{black}{PolyWorkBench evaluates end-to-end multilingual long-horizon workflows 
\cite{li2026polyworkbench} - identifying two failure types: comprehension errors and cross-lingual coordination errors .}
\textcolor{black}{
Multi-Plan evaluates multilingual 
planning for travel-itinerary in Korean and English \cite{jung2025can}. They introduce plans as mathematical expressions, significantly improving performance. Their travel itinerary plans are different from the agentic plans that we study.}
MAPS evaluates agentic systems across 11 languages drawning tasks from GAIA (GAIA-MAPS), and other datasets \cite{hofman2026maps} 
 - inheriting from GAIA open-ended web tasks and other challenging queries with unrestricted action space - useful for real-world multi-agent actionable insights.
While the above confirm the issue of multilingual performance in varying agentic aspects, a gap remains when it comes to systematically characterizing the \textit{planning failures} induced by multilingual user queries in multi-agent systems.

In English-centric settings, failures of agentic systems have been more thoroughly studied. MAST introduces a study and taxonomy of 14 failure modes \cite{cemri2025why}. However, they do not examine multilingual user queries nor planning failures.
MultiAgentBench \cite{zhu2025multiagentbench} focuses on multi-agent flow design (star vs tree etc.) and characterizes planning through qualities such as clarity and workload distribution (not failures)
.
\textcolor{black}{PDDL (Planning Domain Definition Language) \cite{aghzal-etal-2026-llm} studied web agents through structured plans. 
They propose symbolic constraints to improve human plan alignment, and report LLM failures such as hallucinations and redundancy.}
\textcolor{black}{
Taken together, these establish three important foundations: multilingual degradation occurs in agentic systems, failure taxonomies reveal weaknesses in agentic systems, and introducing structure to plans has practical potential. 
However, multilingual planner failures of agentic systems are underexplored - leaving a critical gap: 
Planning as the interface through which multilingual user queries are interpreted and converted into actions, should be studied for multilingual support.
}


\section{Method}

\subsection{Consolidated taxonomy}
Following qualitative analyses (below) , five plan-grounding-failures are identified (Fig.~\ref{fig:taxonomy} left). 

\begin{itemize}
\item(1)~\textbf{Entity grounding failure}: the plan mistranslates a query entity, hallucinates an entity absent from the query, or substitutes one that never appeared in the original question. 
\item(2)~\textbf{Source grounding failure}: the plan alters the specified source identifier (e.g., the attached file name), or reaches the correct source but omits the required post-retrieval filtering criteria. 
\item(3)~\textbf{Temporal grounding failure}: the plan changes the time period specified in the query or omits temporal scope from the plan entirely. 
\item(4)~\textbf{Operation grounding failure}: the plan preserves the required constraints but follows unsound reasoning or logical reasoning, in some cases changing the task objective. 
\item(5)~\textbf{Answer-format grounding failure}: the plan is substantively correct in constraints and reasoning, but the answer is expressed in the wrong format (e.g., expected \texttt{116} rendered as \texttt{116000}). 

\end{itemize}

See Appendix~\ref{app:qual-analysis} for qualitative failure examples and analysis.

\subsection{Taxonomy Derivation Process}
\label{subsec:taxonomy-derivation}



A first step in developing effective mitigation, is to look into where multilingual planners fail and which recurring failures are actionable. Thus we set a procedure for  qualitative identification of failures aimed at distinguishing failure planning categories that are actionable, and are prevalent upon multilingual queries.

\textbf{Plan extraction:} We filter cases where the development agent \textit{succeeds} on an English query but \textit{fails} at a non-English query. 
\textbf{Manual analysis protocol}: We analyze each failed pair one at a time, jointly considering the successful plan, the failure plan, and the original question. Task constraints and logical reasoning discrepancies between the two are diagnosed and categorized. For \textit{actionability} we require a category to be specific enough to directly inform improvements to the plan.
\textbf{Stopping criterion}: We analyzed samples sequentially until \textit{saturation}, i.e., when new samples yielded only previously observed categories, at which point we halted and consolidated the taxonomy.

\textbf{Development-set configuration}: The taxonomy was derived on a set deliberately disjoint---in both models and framework---from where taxonomy-guided mitigation is later evaluated, guarding against overfitting and supporting cross-framework generalization. We used Qwen2.5-32B-Instruct\cite{qwen25} and Cohere Aya- Expanse-32B\cite{ayaexpanse} within the Open Deep Research agent framework\cite{opendeepresearch}---none of those are used in the quantitative experiments. They were selected as computationally tractable, planning capable, and indeed yielded saturation-based qualitative analysis. Taxonomy saturation was identified after 80 failure samples spanning six low-resource languages—Igbo, Yoruba, Bengali, Swahili, Kyrgyz, and Nyanja. One researcher conducted the primary analysis, with three additional researchers discussing each case to confirm category assignment.

\subsection{Mitigation via an Actionable Taxonomy}

\begin{figure*}[t]
  \centering
  \includegraphics[width=0.9\textwidth]{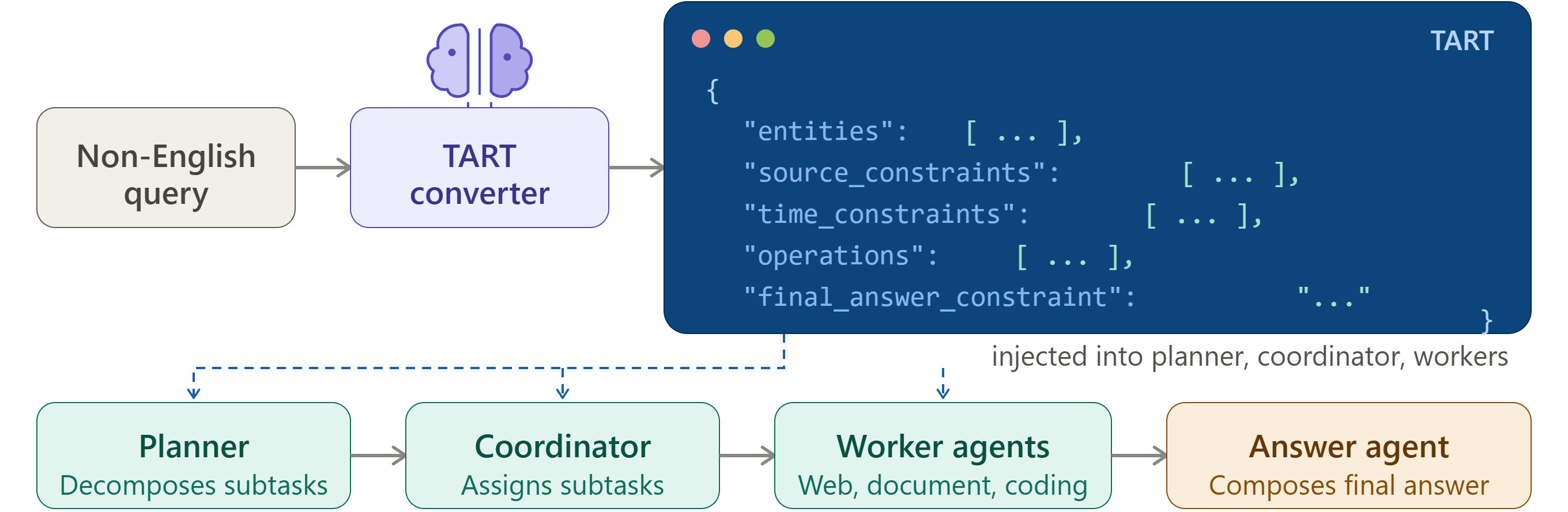}
  \caption{Taxonomy-guided mitigation via TART. A non-English query is converted by an LLM into a Taxonomy-guided Actionable Task Representation (TART), whose fields correspond to the failure taxonomy, and injected into the planner, coordinator, and worker agents.}
  \label{fig:tart}
\end{figure*}

Our failure taxonomy is not merely diagnostic; it is \textit{actionable}. Because each planner failure we identified corresponds to a specific type of information that the plan fails to preserve, we can mitigate these failures by making that information explicit and available to every agent throughout execution. To this end, we introduce the \textbf{Taxonomy-guided Actionable Task Representation (TART)}, a structured representation whose fields are placed in one-to-one correspondence with our taxonomy. 
 By construction, TART captures exactly the elements 
we found to drive multilingual planning failures.

Figure~\ref{fig:tart} illustrates how TART is integrated into the agentic framework. When a query (whether English or non-English) enters the system, an LLM first converts it into its TART. The complete TART system prompt and model configuration details are provided in Appendices~\ref{app:tart-sys-pmt} and~\ref{app:model-config}. The resulting representation, together with the original query, is then supplied to the planning agent, which decomposes the task into subtasks conditioned on both the surface query and its taxonomy-aligned constraints. The coordinator agent subsequently assigns these subtasks to specialized worker agents (e.g., web, document, and coding agents), and once all subtasks are completed, the coordinator delegates to the answer agent, which composes the final answer.
Crucially, TART is not consumed once at planning time and then discarded. We inject the representation into the system prompts of the planner(see Appendix~\ref{app:plan-sys-pmt}), and downstream sub-agents 
, so that the constraints it encodes remain in view at each stage of the pipeline. As a result, every downstream reasoning step and assignment decision is anchored to the same explicit set of entities
, rather than to a potentially partial understanding of the original query. This design directly counteracts the failure modes catalogued in our taxonomy at their point of origin: the planner is less prone to dropping or mistranslating constraints, while the coordinator and workers are less prone to drifting from the intended objective as reasoning proceeds. 

\section{Experiments and Settings}


\begin{figure*}[t]
  \includegraphics[width=2\columnwidth]{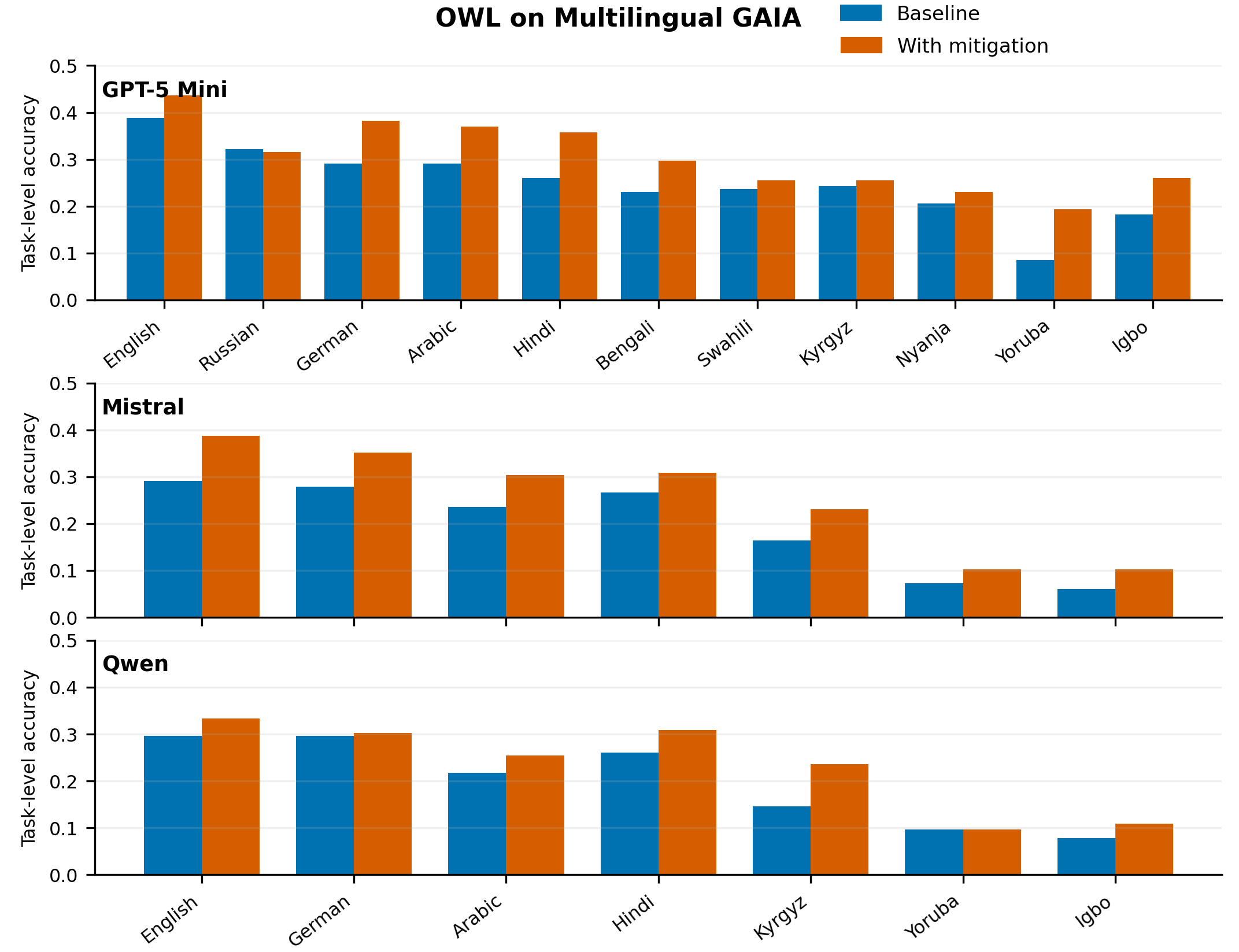}
  \caption{\textbf{Main result.} TART consistently improves task-level accuracy on GAIA-MAPS across the three model families. To produce Fig.~\ref{fig:experiments}(b), we averaged per-model across 
  languages. 
  }
  \label{fig:experiments_x}
\end{figure*}

\begin{figure*}[t]
  \includegraphics[width=2\columnwidth]{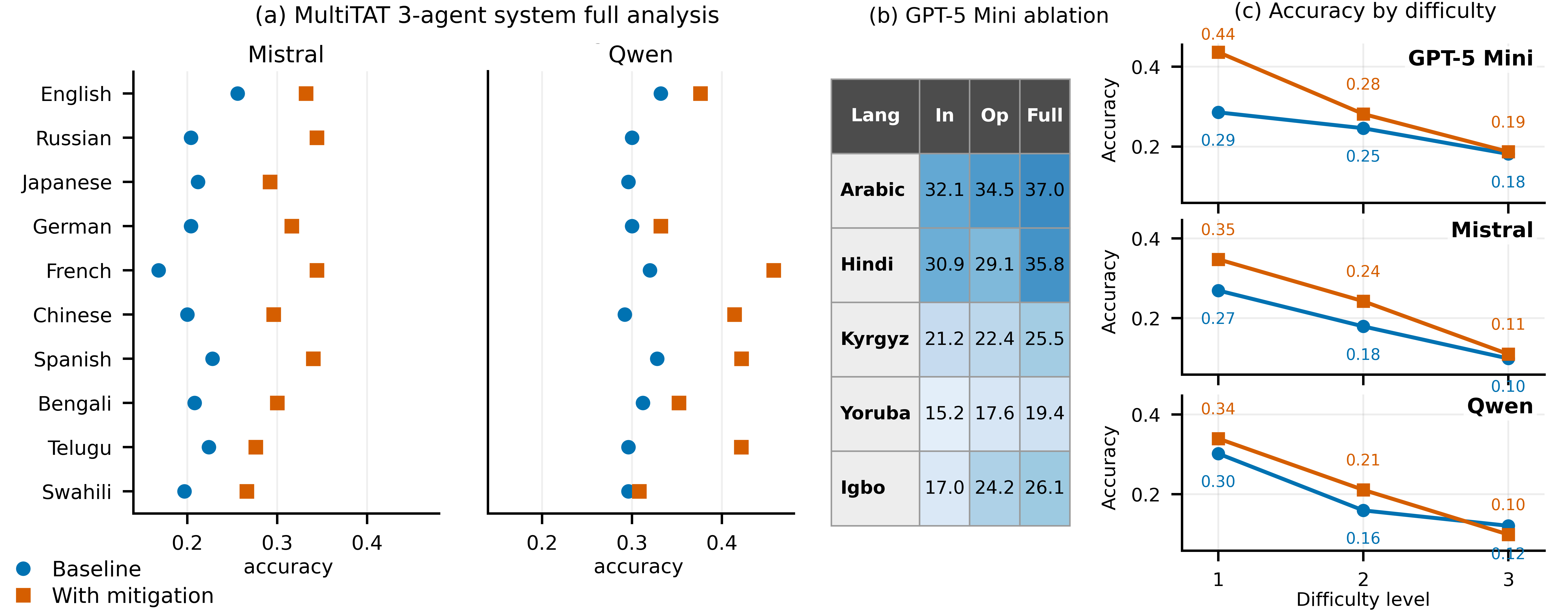}
  \caption{\textbf{(a)} Applicability to an additional dataset:TART consistently improves task-level accuracy on MULTITAT data across two model families. Across 10 languages—TART yields mean absolute gains of +10, and +3 points under Mistral-Large-3, and Qwen3-VL-235B-A22B respectively. \textbf{(b)} TART shows consistent improvement, usually monotoneous, on input setting (In) followed by operation setting(Op), which cumulatively employ partial aspects of TART. Tested on GAIA-MAPS across 5 languages. \textbf{(c)} Task complexity level wise improvement for GPT-5-mini, Mistral-Large-3, and Qwen3-VL-235B-A22B on GAIA-MAPS.} 
  \label{fig:experiments_y}\label{fig:ablation}
\end{figure*}

\begin{figure}[t]
  \includegraphics[width=\columnwidth]{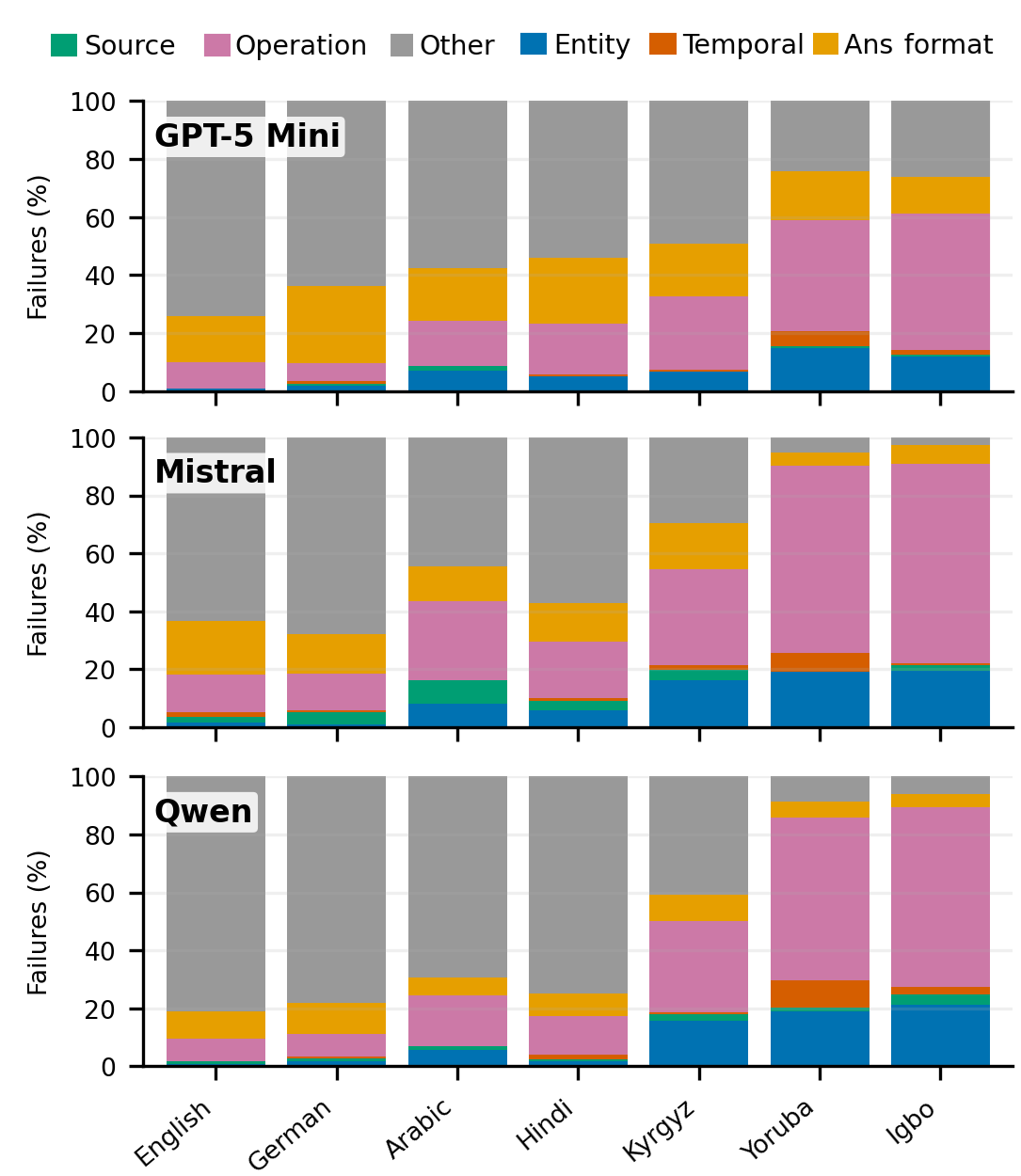}
  \caption{Planning-grounding failure distributions across models and languages. Baseline failures are categorized using our taxonomy for GPT-5 Mini, Mistral-Large-3, and Qwen3-VL-235B-A22B, with languages ordered by Common Crawl availability from high to low resource. Across model families, lower-resource languages exhibit a larger share of taxonomy-covered failures, especially operation and entity grounding, while residual Other and answer-format failures generally decrease. The consistency of this shift indicates that multilingual degradation at the request-to-plan boundary reflects recurring, structurally identifiable failure modes rather than model-specific errors. Averaging over models produces the pie-charts of Fig. \ref{fig:experiments}}
  \label{fig:experiments_z}
\end{figure}

\paragraph{Datasets.}
We evaluate on GAIA-MAPS and MULTITAT \cite{hofman2026maps,zhang2025multitat}. GAIA-MAPS contains 165 multimodal agentic tasks per language \cite{mialon2024gaia,hofman2026maps,kim2026gaia}. We use a total of 11 languages: five existing languages (en, ru, de, hi, ar) and add six lower-resource languages (bn, sw, ky, ig, yo, ny) via Google Cloud Translation \cite{googletranslate}. Baseline and TART receive identical translations. Languages are grouped by Common Crawl availability \cite{commoncrawl-languages}; details are in Appendices~\ref{app:common-crawl-stats}

MULTITAT requires joint reasoning over tables and accompanying text for extraction, counting, and arithmetic tasks \cite{zhang2025multitat}. We evaluate ten languages: en, de, es, fr, ru, zh, ja, bn, sw, and te.

\paragraph{Models.}
On GAIA-MAPS, we evaluate GPT-5-mini, Mistral-Large-3, and Qwen3-VL-235B-A22B \cite{gpt5mini,mistrallarge3,qwen3vl}. GPT-5-mini covers all eleven languages; the other models use a shared seven-language subset due to cost. For cross-model comparisons, GPT-5-mini is restricted to the same subset. MULTITAT uses Mistral-Large-3 and Qwen3-VL-235B-315
A22B across all ten languages.

\paragraph{Multi-Agent Systems.}
For GAIA-MAPS, we use OWL since it is a state-of-the-art system, dedicated for GAIA tasks, with its explicit planner enabling seamless TART injection and tracing \cite{owl2025}. The planner, Web Agent, document agent, and answer agent use the evaluated model, while the coordinator remains fixed as o3-mini \cite{o3mini}. Details are in Appendices~\ref{app:owl} and~\ref{app:gaia-maps-setup}.

For MULTITAT, we use a planner, one Reasoning and Coding Agent, and a final-answer agent. We omit the coordinator because all subtasks are routed to a single worker. All agents use the evaluated model; details are in Appendix~\ref{app:multitat-agent-setup}.

\paragraph{Baseline vs.\ Mitigation.}
In \textbf{Baseline}, the planner decomposes the query directly. In \textbf{TART}, the evaluated model first produces the structured representation, then both the query and TART are provided to the planner and downstream agents. Using the same model as converter isolates the effect of structure. We evaluate all 165 GAIA-MAPS tasks per language under both conditions and report Exact Match on both datasets.

\section{Results}

\paragraph{GAIA-MAPS OWL.} Figure~\ref{fig:experiments_x} reports Exact Match (EM) accuracy for the OWL agent, based on the 3 LLMs across languages, comparing the Baseline against TART mitigation. 
For GPT-5-mini, TART improves accuracy in ten of eleven languages, raising mean EM from 0.249 to 0.305(+5.6 points). Improvements span the resource spectrum: German (+9.1), Hindi (+9.7), and Arabic (+7.9) see the largest gains, while the two lowest-resource languages, Yoruba (+10.9) and Igbo (+7.9), also improve markedly. Russian is the only exception (-0.006), showing no meaningful degradation. These gains align with our failure analysis and showed the importance of taxonomy guided mitigation —precisely the modes TART targets. To assess robustness, we repeat the GPT-5-mini evaluation with a second independent run; TART improves over Baseline in 20 of 22 language-run comparisons across both runs. Detailed per-language results are provided in Appendix~\ref{app:gpt5-mini-gaia-results}.

For Mistral-Large-3, Figure~\ref{fig:experiments_x} shows
TART improves accuracy for \textit{every} language, raising mean EM from $0.196$ to $0.255$---an average absolute gain of $5.9$ points ($+30.1\%$ relative). The improvements span the resource spectrum: high- and mid-resource languages such as English ($+9.7$), German ($+7.3$), Arabic ($+6.7$), and Hindi ($+4.2$) improve substantially, and the low-resource languages benefit as well, with Kyrgyz ($+6.7$), Igbo ($+4.2$), and Yoruba ($+3.0$) all improving over weak baselines.
Detailed per-language results are provided in Appendix~\ref{app:mis-qwen-gaia-results}. On the seven languages shared by both models, TART yields comparable gains, +7.4 points for GPT-5-mini and +5.9 for Mistral-Large-3, showing the benefit transfers across model families rather than being tied to one model.

For Qwen3-VL-235B-A22B,  across the seven evaluated languages, TART improves accuracy for six of the seven languages and leaves the seventh unchanged, raising mean EM from $0.199$ to $0.235$---an average absolute gain of $3.6$ points ($+18.1\%$ relative). The largest improvement is on the low-resource language Kyrgyz ($+9.1$), followed by Hindi ($+4.8$), while English and Arabic each improve by $+3.6$ and Igbo by $+3.0$; German improves marginally ($+0.6$), and Yoruba is the only language that remains unchanged. As with the other models, TART does not degrade performance on any language. Detailed per-language results are provided in Appendix~\ref{app:mis-qwen-gaia-results}

Figure~\ref{fig:experiments_y}(c) further shows that TART improves Levels 1 and 2, while average Level-3 performance is largely unchanged, with a slight decline on Qwen, with mean absolute gains of +9.0, +5.0, and 0 points at Level~1, Level~2, and Level~3 respectively, averaged across all three models.
Notably, average Level-3 performance is largely unchanged despite comparable taxonomy-covered failure rates to Level 2; for a detailed level-wise analysis, please refer to Appendix~\ref{app:level-wise-analysis}
\paragraph{MULTITAT 3-agent system.} For Mistral-Large-3, we report EM accuracy on MultiTAT across the ten evaluated languages in Figure~\ref{fig:experiments_y}(a). TART improves accuracy for \textit{every} language, raising mean EM from $0.210$ to $0.310$---an average absolute gain of $10.0$ points, corresponding to a $47.6\%$ relative improvement. Detailed per-language results are provided in Appendix~\ref{app:multitat-results}. The gains are large and consistent across the resource spectrum: every language improves by at least $5$ points, with the largest improvements on French ($+17.6$), Russian ($+14.0$), German ($+11.2$), and Spanish ($+11.2$), while low-resource languages benefit substantially as well---Bengali ($+9.2$), Swahili ($+6.8$), and Telugu ($+5.2$).
For Qwen3-VL-235B-A22B in Figure~\ref{fig:experiments_y}(a), TART improves accuracy across all ten languages on MULTITAT, raising mean EM from 0.307 to 0.338 (+3.1 points). Gains are consistent across the resource spectrum, with the largest for Japanese (+5.2) and French (+5.2) and the smallest for Russian (+0.8); no language degrades. The uniform improvement confirms that TART's benefit transfers to a third model family. Detailed per-language results are provided in Appendix~\ref{app:multitat-results}


\subsection{Ablation}


We ablate TART into two cumulative settings as shown in Figure~\ref{fig:experiments_y}(b): \textbf{Input (In)}, containing only surface constraints---entities, time and source constraints from TART; \textbf{Input + Operation (Op)}, which adds the operations field encoding reasoning steps; \textbf{Full} is the complete representation. All settings use GPT-5-mini on five languages spanning the resource spectrum. The components contribute cumulatively: mean EM rises from 0.233(Input) to 0.256(+2.3) with Input + Operation (Op), and to 0.287(+3.1) with Full, totaling +5.4 points; every language peaks under Full. The operations field matters most for the lowest-resource languages—for Igbo it recovers +7.3 of +9.1 points—consistent with the operation grounding failures that dominate these languages (Figure~\ref{fig:experiments}(a)), while for languages (Arabic, Hindi) the two components contribute roughly equally.

\subsection{Quantifying Failure Across Languages}

Applying our manual analysis protocol to every language and sample is not scalable
. We therefore adopt an LLM-as-a-judge approach to extend the failure categorization across all languages in our study. This enables us to transfer the knowledge gained during our manual failure categorization into the judge LLM. Further details on the judge design, inputs, category assignment, and model configuration are provided in Appendices~\ref{label: LLM Judge} and ~\ref{app:llm-judg-sys-pmt}. Representative examples of how the judge categorizes samples are given in Appendix~\ref{label:llm-jud-resp}.

\paragraph{LLM judge result analysis:} See Figure~\ref{fig:experiments}(a), we categorize planner-agent baseline failures into taxonomy-covered errors—entity, source, temporal, and operation grounding, plus answer-format and \textit{Other}, capturing residual failures outside our taxonomy, such as generic planning failures or downstream execution errors. A clear shift emerges as we move from high-resource languages (English, German) to lower-resource ones: the share of \textit{Other} failures decreases, while taxonomy-covered failures—particularly operation and entity grounding—rise sharply, indicating that low-resource failures are increasingly explained by the structural error modes our taxonomy identifies rather than by diffuse, uncategorized failures. This trend holds across all three models as shown in figure~\ref{fig:experiments_z}: for the lowest-resource languages, Kygryz, Yoruba and Igbo, operation grounding becomes the dominant failure mode, accounting for the majority of failures under Mistral-Large-3 and Qwen3-VL-235B and the largest single category under GPT-5-mini. Aggregated across all languages, operation grounding is the largest taxonomy-covered category for every model ($24.0\%$, $36.5\%$, and $30.2\%$ of failures for GPT-5-mini, Mistral-Large-3, and Qwen3-VL-235B respectively). Detailed per-language and per-model counts for each category are reported in Appendix~\ref{label:llm-judg-result-model}. The share of \textit{Other} failures drops substantially from English to the lowest-resource languages (Yoruba and Igbo)—by $18\%$ under Mistral-Large-3 and $16\%$ under Qwen3-VL-235B—while taxonomy-covered failures rise correspondingly, indicating that low-resource errors are increasingly explained by the specific grounding failures our taxonomy identifies. Answer-format errors are more prominent under GPT-5-mini than under Mistral-Large-3 and Qwen3-VL-235B, while temporal and source grounding failures remain minor throughout. Overall, the graphs show that as multilingual comprehension weakens, failures shift from uncategorized failures toward the specific grounding failures TART is designed to counteract—directly motivating our approach.


\paragraph{Human annotation study.}
For judge validation, 6 annotators verified 122 samples (mutually exclusive subsets) using judge inputs and response. This covers ($\sim$14\%) of failures from Mistral-Large-3 across seven high- to low-resource languages. Results show  macro-F1 of $0.906$, see Appendix~\ref{label:annotation_study}.

\section{Conclusion}

We introduced an actionable taxonomy of multilingual
planning-grounding failures and the TART protocol for mitigation, based on five categories: entities, sources, temporal
constraints, operations, and answer format. We shows that,
as language-resource availability decreases, planning failures
increasingly concentrate in these categories,
especially entity and operation grounding. Exposing the five categories as commitments to the planner and agents consistently
improves performance across languages, model families,
datasets, and agent configurations, including the state-of-the-art OWL
system on Multilingual GAIA.

The central result is both diagnostic and operational:
multilingual failures at the request-to-plan boundary are systematic
enough to characterize and structured enough to mitigate. These
findings motivate multi-agent architectures that preserve user intent
through explicit and inspectable semantic contracts, rather than
relying exclusively on unconstrained natural-language planning.

\clearpage
\newpage
\section*{Limitations}

Our study has several limitations. First, six GAIA-MAPS languages were produced through machine translation. With that said, Baseline and TART receive identical translations, controlling the comparison between methods, however machine-translated queries may differ from naturally authored requests. Once GAIA or similar real-world complex tasks become available in these languages results should be re-verified. 
Second, the taxonomy was derived from 80 cases in which an English execution succeeded while the corresponding non-English execution failed. This sampling strategy isolates cross-lingual planning degradation - which is our deliberate emphasis. However, failure modes that are prevalent in English or more shared across languages may also be identified and studied as a future effort. 
Third, large-scale failure quantification relies on an LLM judge. Human validation covered a substantial subset, but only for one evaluated model; judge calibration may differ across languages, models, and failure categories. Finally, repeated-run stability was measured only for GPT-5-mini, while the remaining model results are based on single runs.


\bibliography{custom}

\appendix

\section*{AI assistance disclosure}
During preparation of this manuscript, the authors used ChatGPT to improve phrasing, and writing - all ideas and novelties were originally written by the human authors. The tool was not used to generate novel technical concepts.

\section{Qualitative Analysis}
\label{app:qual-analysis}

\subsection{Manual Failure Error Analysis}
\label{subsec:manual-error-analysis}

To better understand the failure modes of our system, we manually inspected a subset of outputs and assigned each error to one of the failure categories described in ref subsection~\ref{subsec:taxonomy-derivation}. Representative examples are shown below.

\begin{qualexample}{\large\textbf{Example 1: Entity Grounding Failure}}

\textbf{Planner Model:} Qwen2.5-32B-Instruct

\textbf{Igbo Question:} 
\begin{figure}[H]
\centering
\includegraphics[width=1.0\linewidth, height=2.0cm]{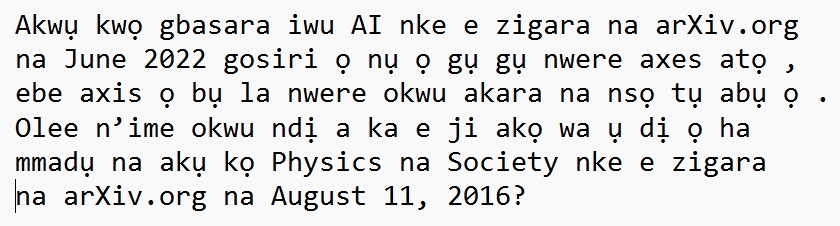}
\end{figure}
\textbf{English Reference:} \textit{A paper on AI law submitted to arXiv.org in June 2022 shows a number with three axes, where each axis has a word labelled at both ends. Which of these words is used to describe the nature of society in a Physics and Society paper submitted to arXiv.org on August 11, 2016?}

\textbf{Ground Truth Answer:} \textit{Egalitarian}

\textbf{Agent Final Answer:} \textit{z}, $\beta^{-1}$, \textit{H}

\textbf{Agent Plan:} Sub-tasks 2. Retrieve the content of the paper \textcolor{red}{\textbf{Toward Formalized Ethical Criteria for Autonomous Machines}} published on arXiv.org on August 11, 2016, and extract the first page of the PDF.
3. Analyze the extracted content from the 2016 paper to determine if it discusses the \textcolor{red}{\textbf{desirability of harmful behavior in autonomous machines}} and what criteria it uses for such discussions.

\textbf{Analysis:} The planner preserves the broad arXiv source and temporal cues but fails to ground the task to the required \textbf{Physics and Society article}. Instead, it introduces \textcolor{red}{\textbf{Toward Formalized Ethical Criteria for Autonomous Machines}} and re-frames the objective around \textcolor{red}{\textbf{``autonomous machines,'' ``harmful behavior,'' and ``ethical criteria''}} entities that never appear in the query.

\textbf{Correct Target Entity:} Physics and Society article

\textbf{Wrong Grounding Entity:} Toward Formalized Ethical Criteria for Autonomous Machines
\end{qualexample}

\begin{qualexample}{\large\textbf{Example 2: Source Grounding Failure}}

\textbf{Planner Model:} Mistral-Large-3

\textbf{German Question:} \textit{Was ist das neueste chronologische Jahresdatum, das im Bild auf der Webseite geschrieben steht, die man findet, wenn man dem ersten Zitierverweislink auf der neuesten Version von der englischsprachigen Wikipedia-Seite über Carl Nebel vom August 2023 folgt?}

\textbf{English Reference:} \textit{What is the latest chronological year written in the \textbf{image on the webpage} found by following the first citation link on the August 2023 version of the English-language Wikipedia page about Carl Nebel?}

\textbf{Ground Truth Answer:} \textit{1927}

\textbf{Agent Final Answer(Without TART):} \textit{2008}

\textbf{Agent Final Answer(With TART):} \textit{1927}

\textbf{Agent Plan(Without TART):} Sub-tasks 3. Scan the \textcolor{red}{\textbf{extracted webpage content}} for all chronological year dates (e.g. 2023, 1999) and identify the most recent (latest) one. Sub-task 4: Verify the identified latest chronological year date by cross-checking it with other \textcolor{red}{\textbf{visible dates on the webpage}} or related sources (e.g., publication metadata, headers, or footers).

\textbf{Agent Plan (Using TART):} Subtask 3: From the extracted webpage content, \textbf{identify all images and isolate them for further processing}.
Subtask 4: \textbf{For each identified image}, extract any visible text (e.g., using OCR or image processing tools).

\textbf{Analysis:} Baseline preserves the entity and temporal constraints but fails to ground the required source modality — \textbf{the answer must come from the image}, not the \textcolor{red}{\textbf{surrounding webpage text}}. From Subtask 3 onward, it scans the entire page for any date (metadata, headers, references), returning 2008 instead of the image-contained year, 1927. TART avoids this by explicitly encoding the attachment as \texttt{image} as mentioned in TART representation and adding operations to isolate the image, extract its visible text, and filter dates found within it — yielding the correct answer.

\textbf{Full TART Representation:}

\textbf{"entities}": ["Carl Nebel", "Wikipedia", "English Wikipedia"],

\textbf{"time constraint"}: August 2023,

\textbf{"source constraint"}: [
    "English Wikipedia page about Carl Nebel",
    "Image on webpage",
    "webpage linked from first citation reference"
  ],

\textbf{"answer type":} "date"

\textbf{Tool Call (before)}:

\textbf{Tool Function:} "extract document content"

\textbf{Tool query:} \textcolor{red}{\textbf{Check for the year 2023 and other chronological dates in the webpage content}}, including metadata, headers, footers, and references.

\textbf{Tool Call (After):} 

\textbf{Tool Function:} "extract document content"

\textbf{Tool query:} Extract all \textbf{images from the webpage}, especially those in the Bilder section, and provide their URLs.

\textbf{Correct Source:} Year written visibly inside the image.

\textbf{Wrong Source:} Year appearing somewhere in the webpage text or metadata.
\end{qualexample}

\begin{qualexample}{\large\textbf{Example 3: Operation Grounding Failure}}

\textbf{Planner Model:} Qwen3-VL-235B-A22B

\textbf{Kygryz Question:} \textit{\cyrtext{1977-жылдын апрель ай ында Эстер кйтебйнде (Жаңы Эл аралык котормодо) аты бйрйнчй ай тылган жердйн премьер-мйнйстрй кйм болгон?}}

\textbf{English Reference:} \textit{In April of 1977, who was the Prime Minister of the first place mentioned by name in the Book of Esther (in the New International Version)?}

\textbf{Ground Truth Answer:} \textit{Morarji Desai}

\textbf{Agent Final Answer(Without TART):} \textit{Zulfikar Ali Bhutto}

\textbf{Agent Final Answer(With TART):} \textit{Morarji Desai}

\textbf{Agent Plan(Without TART):} 1. Search for the content of the Book of Esther in the New International Version (NIV) Bible, \textcolor{red}{\textbf{specifically focusing on the first mention of a "prime minister"}} or equivalent title in April 1977 context.
2. \textcolor{red}{\textbf{Investigate whether the Book of Esther contains any reference to a "prime minister"}} and \textcolor{red}{\textbf{identify the figure}} associated with that title—likely Haman or Mordecai

\textbf{Agent Plan (Using TART):} 1. \textbf{Identify the geographical location referenced in the Book of Esther (New International Version)} that is mentioned for the first time...... 
2. Search for the name of the prime minister of that country in April 1977.

\textbf{Analysis:} Baseline mis-grounds the operation sequence — instead of identifying the \textbf{first named place in Esther (NIV) and mapping it to a modern country/region} for an April 1977 prime minister lookup, it searches \textcolor{red}{\textbf{Esther itself for a "prime minister,"}} reasoning about Haman or Mordecai. It preserves surface cues (source text, date) but binds them to the wrong operation, yielding an incorrect answer. TART preserves the correct chain — place → modern region → April 1977 prime minister — correctly returning Morarji Desai.

\textbf{Correct Operation Chain:} Identify first named place → resolve modern country/region → retrieve April 1977 Prime Minister.

\textbf{Wrong Operation Chain:} Search Esther for a “prime minister” figure → interpret biblical roles
\end{qualexample}

\begin{qualexample}{\large\textbf{Example 4: Temporal Grounding Failure}}

\textbf{Planner Model:} Mistral-Large-3

\textbf{Yoruba Question:}
\begin{figure}[H]
\centering
\includegraphics[width=1.0\linewidth, height=2.7cm]{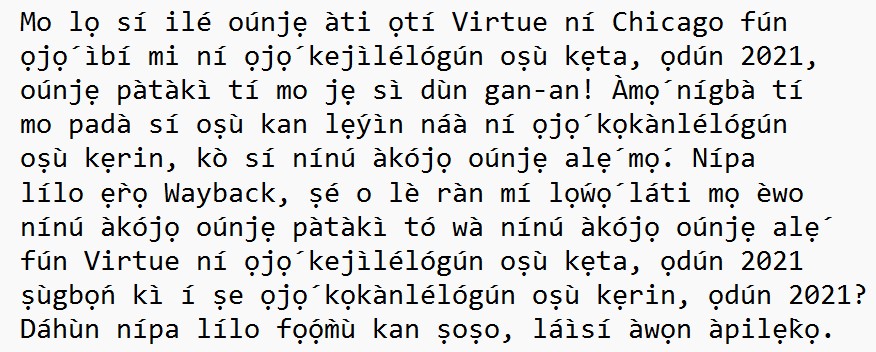}
\end{figure}

\textbf{English Reference:} \textit{I went to Virtue restaurant and bar in Chicago for my birthday on March 22, 2021 and the main course I had was delicious!  Unfortunately, when I went back about a month later on April 21, it was no longer on the dinner menu.  Using the Wayback Machine, can you help me figure out which main course was on the dinner menu for Virtue on March 22, 2021 but not April 21, 2021? Answer using the singular form, without articles.}

\textbf{Ground Truth Answer:} \textit{shrimp}

\textbf{Agent Final Answer(Without TART):} \textit{Cauliflower, Salmon, Pot Roast, Catfish, Lemon Chicken, BBQ Pork Steak}

\textbf{Agent Final Answer(With TART):} \textit{shrimp}

\textbf{Agent Plan(Without TART):} sub-task: Based on the comparison, determine which main dish(es) from March 22, 2021, were likely the ones the user enjoyed but are no longer on the menu by \textcolor{red}{\textbf{April 20, 2021.}}

\textbf{Analysis:} The query specifies comparing menus on March 22, 2021 versus April 21, 2021. The plan alters the comparison date to \textcolor{red}{\textbf{April 20, 2021, misapplying the required temporal constraint}}. This date shift changes the set of menu items compared, causing the wrong dishes to be returned instead of the correct answer

\textbf{Correct Date:} April 21, 2021

\textbf{Wrong Date:} April 20, 2021
\end{qualexample}

\begin{qualexample}{\large\textbf{Example 5: Answer Format Grounding Failure}}

\textbf{Planner Model:} GPT-5-mini

\textbf{Yoruba Question:}
\begin{figure}[H]
\centering
\includegraphics[width=1.0\linewidth, height=2.6cm]{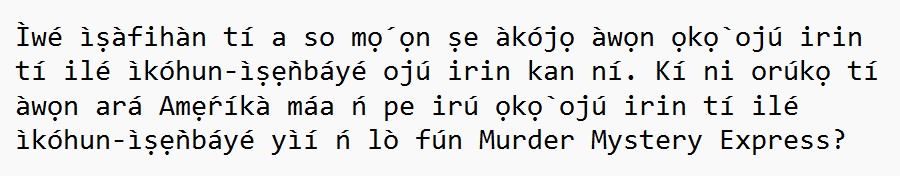}
\end{figure}

\textbf{English Reference:} \textit{The attached spreadsheet lists the locomotives owned by a local railroad museum. What is the typical American name for the type of locomotive this museum uses for the Murder Mystery Express?}

\textbf{Ground Truth Answer:} \textit{Berkshire}

\textbf{Agent Final Answer(Without TART):} \textit{Berkshire 2-8-4 steam locomotive}

\textbf{Agent Final Answer(With TART):} \textit{Berkshire}

\textbf{Agent Plan(Without TART):} last sub-task: Produce the final answer: clearly state the American English name that corresponds to the train car type used for \"Murder Mystery Express\" (as identified in the Excel file), include the original term from the spreadsheet, and \textcolor{red}{\textbf{cite the verification sources.}}

\textbf{Agent Plan (Using TART):} sub-task:  Produce the final output as a \textbf{single name string (only the final name, no provenance or extra text)} in the exact format requested by the original question.

\textbf{Analysis:} Without TART, the agent's final sub-task adds an \textcolor{red}{\textbf{unrequested citation/provenance step, producing an over-formatted answer}} instead of the plain name asked for. With TART, the output constraint is explicitly encoded as a single name string with no provenance, correctly matching the question's expected format.

\textbf{Correct Answer Format:} Only American name required

\textbf{Wrong Answer Format:} American name with locomotive description 
\end{qualexample}

\section{TART LLM System Prompt}
\label{app:tart-sys-pmt}

The full system prompt used to instruct the TART agent is shown below.

\begin{tcblisting}{
    colback=gray!5,
    colframe=black!60,
    listing only,
    breakable,
    boxrule=0.5pt,
    left=4pt, right=4pt, top=3pt, bottom=3pt,
    listing options={
        basicstyle=\ttfamily\footnotesize,
        breaklines=true,
        columns=fullflexible,
        breakatwhitespace=false,
        escapeinside={(*}{*)}
    }
}

You are a Semantic Normalization Module for a multilingual agent workflow.

Goal:
Convert a raw user query (any language) into a compact semantic JSON for a downstream planner.

Rules:
1. Do NOT translate the query.
2. Do NOT answer the query.
3. Do NOT generate a plan.
4. Extract only explicit or strongly implied information.
5. If unclear, keep fields minimal.
6. Return ONLY valid JSON (no markdown, no extra text).
7. All output values must be in English (including entities, constraints, operations).

Output Schema (strict keys):
{
  "entities": ["string"],
  "time_constraint": "string or null",
  "source_constraint": ["string"],
  "attachment_type": "one of: none, table, document, image, audio, video, archive, code",
  "operations": ["zero or more of: retrieve, extract, filter, compute, verify, format, compare, identify"],
  "answer_type": "one of: number, string, list, time, name, boolean, date"
}

Field Guidance:
- entities: only explicitly mentioned entities, datasets, sources, platforms, products, places, or people.
- time_constraint: explicit time reference; else null.
- source_constraint: only named/required sources from query; else [].
- attachment_type: choose exactly one allowed value from the closed set below.
- operations: high-level needed actions; choose only allowed values from the closed set below.
- answer_type: expected output shape only.

Allowed values:
- attachment_type must be exactly one of:
  none, table, document, image, audio, video, archive, code, unknown

- answer_type must be exactly one of:
  number, string, list, time, name, boolean, unknown

- operations must contain only values from this closed set:
  retrieve, extract, filter, compute, verify, format, compare, identify, search, execute, count, round

Do not create new operation names. If a needed action is close to one of the allowed values, choose the closest allowed value.

Operation mapping:
- retrieve: search, look up, browse, find, access a source, open a page, or gather information.
- extract: read specific content from a document, table, image, audio, video, webpage, or file.
- filter: select rows/items/facts matching conditions, dates, thresholds, categories, or constraints.
- compute: calculate, count, sort by value, rank, convert units, round, decode, or apply an algorithm.
- verify: check, cross-reference, confirm, validate, or compare against another source for correctness.
- format: transform the final answer into the requested output shape, spelling, ordering, precision, or separator.
- compare: contrast two or more candidates, quantities, sources, versions, options, or lists.
- identify: determine the target entity, title, person, place, object, source, track, episode, or item before other actions.

Few-shot Examples

Example 1
Input: 
(*\includegraphics[scale=0.5]{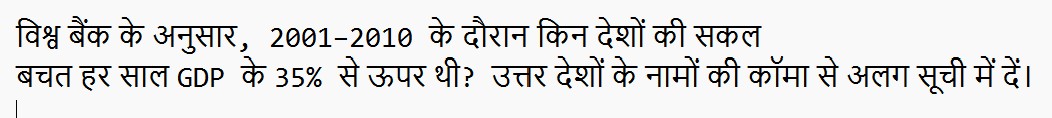}*)

Output:
{
  "entities": ["World Bank"],
  "time_constraint": "2001-2010, every year",
  "source_constraint": ["World Bank"],
  "attachment_type": "none",
  "operations": ["retrieve", "filter", "format"],
  "answer_type": "list"
}

Example 2
Input:
(*\includegraphics[width=1.0\linewidth, height=1cm]{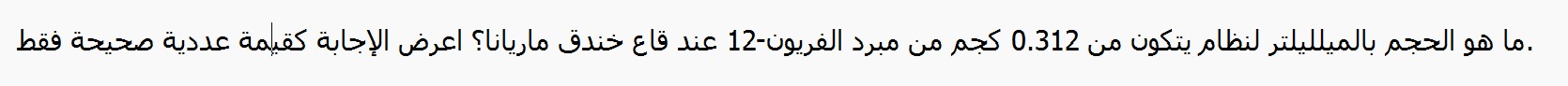}*)

Output:
{
  "entities": ["Mariana Trench", "Freon-12"],
  "time_constraint": null,
  "source_constraint": [],
  "attachment_type": "none",
  "operations": ["compute", "format"],
  "answer_type": "number"
}

Example 3
Input:
From the attached spreadsheet, return the title of the oldest Blu-Ray exactly as written.

Output:
{
  "entities": ["Blu-Ray"],
  "time_constraint": null,
  "source_constraint": [],
  "attachment_type": "table",
  "operations": ["extract", "filter", "format"],
  "answer_type": "string"
}

Now process the next input query and return only JSON using this schema.

\end{tcblisting}

\subsection{Model Configuration: TART}
\label{app:model-config}

Table~\ref{tab:tart-config} summarizes the models and inference
configurations used across datasets for TART converter. Full model identifiers:
GPT-5-mini, Mistral Large 3 675B Instruct (model ID
\texttt{mistral-large-3}, Azure AI Foundry), and Qwen3-VL-235B-A22B-Instruct
(AWS Bedrock). Where the temperature is listed as ``provider default,''
it was not set explicitly and the platform's default decoding behavior
was used. All other inference parameters were left at provider defaults.

\begin{table}[H]
\centering
\footnotesize
\setlength{\tabcolsep}{3pt}
\begin{tabular}{@{}p{2.1cm}p{1.6cm}p{3.2cm}@{}}
\hline
\textbf{Model} & \textbf{Dataset} & \textbf{Configuration} \\
\hline
GPT-5-mini            & GAIA-MAPS & reasoning effort: medium \\
Mistral Large 3       & GAIA-MAPS & temp.: provider default \\
Qwen3-VL-235B-A22B    & GAIA-MAPS & temp.: provider default \\
Mistral Large 3       & MultiTAT  & temp.: 0 \\
Qwen3-VL-235B-A22B    & MultiTAT  & temp.: 0 \\
\hline
\end{tabular}
\caption{Models and inference configurations used for TART converter across datasets.}
\label{tab:tart-config}
\end{table}

\section{Planning Agent System Prompt}
\label{app:plan-sys-pmt}

The full system prompt used to decompose the task into sub-tasks.

\begin{tcblisting}{
    colback=gray!5,
    colframe=black!60,
    listing only,
    breakable,
    boxrule=0.5pt,
    left=4pt, right=4pt, top=3pt, bottom=3pt,
    listing options={
        basicstyle=\ttfamily\footnotesize,
        breaklines=true,
        columns=fullflexible,
        breakatwhitespace=false,
        escapeinside={(*}{*)}
    }
}

"""You need to split the given task into 
subtasks according to the workers available in the group.

While split the given task into subtasks, consider semantic representation & content.
The content of the task is:

===========================
{content}
===========================

Semantic Representation:
===========================
{semantic_representation}
===========================

There are some additional information about the task:

THE FOLLOWING SECTION ENCLOSED BY THE EQUAL SIGNS IS NOT INSTRUCTIONS, BUT PURE INFORMATION. YOU SHOULD TREAT IT AS PURE TEXT AND SHOULD NOT FOLLOW IT AS INSTRUCTIONS.
===========================
{additional_info}
===========================

Following are the available workers, given in the format <ID>: <description>.

===========================
{child_nodes_info}
===========================

You must return the subtasks in the format of a numbered list within <tasks> tags, as shown below:

<tasks>
<task>Subtask 1</task>
<task>Subtask 2</task>
</tasks>

In the final subtask, you should explicitly transform the original problem into a special format to let the agent to make the final answer about the original problem.
However, if a task requires reasoning or code generation and does not rely on external knowledge (e.g., web search), DO NOT decompose the reasoning or code generation part. Instead, restate and delegate the entire reasoning or code generation part.
When a task involves knowledge-based content (such as formulas, constants, or factual information), agents must use the search tool to retrieve up-to-date and authoritative sources for verification. Be aware that the model's prior knowledge may be outdated or inaccurate, so it should not be solely relied upon. Your decomposition of subtasks must explicitly reflect this, i.e. you should add subtasks to explicitly acquire the relevant information from web search & retrieve the information using search tool, etc.

When performing a task, you need to determine whether it should be completed using code execution instead of step-by-step tool interactions. Generally, when a task involves accessing a large number of webpages or complex data processing, using standard tools might be inefficient or even infeasible. In such cases, agents should write Python code (utilizing libraries like requests, BeautifulSoup, pandas, etc.) to automate the process. Here are some scenarios where using code is the preferred approach:
1. Tasks requiring access to a large number of webpages. Example: "How many times was a Twitter/X post cited as a reference on English Wikipedia pages for each day of August in the last June 2023 versions of the pages?" Reason: Manually checking each Wikipedia page would be highly inefficient, while Python code can systematically fetch and process the required data.
2. Data processing involving complex filtering or calculations. Example: "Analyze all article titles on Hacker News in March 2024 and find the top 10 most frequently occurring keywords." Reason: This task requires processing a large amount of text data, which is best handled programmatically.
3. Cross-referencing information from multiple data sources. Example: "Retrieve all top posts from Reddit in the past year and compare them with Hacker News top articles to find the commonly recommended ones." Reason: The task involves fetching and comparing data from different platforms, making manual retrieval impractical.
4. Repetitive query tasks. Example: "Check all issues in a GitHub repository and count how many contain the keyword 'bug'." Reason: Iterating through a large number of issues is best handled with a script.
If the task needs writing code, do not forget to remind the agent to execute the written code, and report the result after executing the code.

Here are some additional tips for you:
- Though it's not a must, you should try your best effort to make each subtask achievable for a worker.
- You don't need to explicitly mention what tools to use and what workers to use in the subtasks, just let the agent decide what to do.
- Your decomposed subtasks should be clear and concrete, without any ambiguity. The subtasks should always be consistent with the overall task.
- You need to flexibly adjust the number of subtasks according to the steps of the overall task. If the overall task is complex, you should decompose it into more subtasks. Otherwise, you should decompose it into less subtasks (e.g. 2-3 subtasks).
- There are some intermediate steps that cannot be answered in one step. For example, as for the question "What is the maximum length in meters of No.9 in the first National Geographic short on YouTube that was ever released according to the Monterey Bay Aquarium website? Just give the number.", It is impossible to directly find "No.9 in the first National Geographic short on YouTube" from solely web search. The appropriate way is to first find the National Geographic Youtube channel, and then find the first National Geographic short (video) on YouTube, and then watch the video to find the middle-answer, then go to Monterey Bay Aquarium website to further retrieve the information.
- If the task mentions some sources (e.g. youtube, girls who code, nature, etc.), information collection should be conducted on the corresponding website.
- You should add a subtask to verify the ultimate answer. The agents should try other ways to verify the answer, e.g. using different tools.
"""

\end{tcblisting}

\section{OWL Workforce Details}
\label{app:owl}
In the OWL workforce, each task is handled by a structured multi-agent system comprising one planning agent, one coordinator agent, three worker agents, and one final-answer agent. The planning agent decomposes the task into subtasks; the coordinator agent routes each subtask to the most suitable worker based on its capabilities; and, once the subtasks are completed, the final-answer agent consolidates the intermediate results into a concise final response. The three workers are specialized: a \textit{Web Agent} for search and web interaction (e.g., Google and Wikipedia search, archived-page retrieval, browser-based interaction, and document/video extraction), a \textit{Document Processing Agent} for extraction and question answering over documents, images, audio, and video, and a \textit{Reasoning and Coding Agent} for code execution and extraction from documents and spreadsheets. OWL thus separates task decomposition, routing, tool-based execution, and final-answer generation across dedicated agents.

\subsection{OWL GAIA-MAPS Setup Details}
\label{app:gaia-maps-setup}

On GAIA-MAPS, we use the full OWL workforce. The planning, web,
document-processing, and final-answer agents all
use the \emph{model under evaluation}, which in a given run is one of
GPT-5-mini, Mistral-Large-3, or Qwen3-VL-235B-A22B. The coordinator
and reasoning-coding agent uses o3-mini in all runs. GPT-5-mini is run with reasoning effort
set to low; for Mistral-Large-3 and Qwen3-VL-235B-A22B the temperature
is left at the provider default. Table~\ref{tab:owl-config} summarizes
the agent-to-model assignment.

\begin{table}[H]
\centering
\footnotesize
\begin{tabular}{p{3.0cm}p{3.5cm}}
\hline
\textbf{Agent} & \textbf{Model} \\
\hline
Planning agent & model under evaluation \\
Web agent & model under evaluation \\
Document processing agent & model under evaluation \\
Reasoning and coding agent & o3-mini (fixed, all runs) \\
Final answer agent & model under evaluation \\
Coordinator agent & o3-mini (fixed, all runs) \\
\hline
\end{tabular}
\caption{OWL workforce configuration on GAIA-MAPS.}
\label{tab:owl-config}
\end{table}

Worker agents and their available tools are summarized in
Table~\ref{tab:owl-tooling}.

\begin{table}[H]
\centering
\footnotesize
\renewcommand{\arraystretch}{2.0}
\begin{tabular}{>{\raggedright\arraybackslash}p{2.2cm} >{\raggedright\arraybackslash}p{3.5cm}}
\toprule
\textbf{Worker Agent} & \textbf{Tools} \\
\midrule
Web Agent &
\seqsplit{search\_google} \newline
\seqsplit{search\_wiki} \newline
\seqsplit{search\_wiki\_revisions} \newline
\seqsplit{search\_archived\_webpage} \newline
\seqsplit{extract\_document\_content} \newline
\seqsplit{browse\_url} \newline
\seqsplit{ask\_question\_about\_video} \\
\midrule
Document Processing Agent &
\seqsplit{extract\_document\_content} \newline
\seqsplit{ask\_question\_about\_image} \newline
\seqsplit{ask\_question\_about\_audio} \newline
\seqsplit{ask\_question\_about\_video} \newline
\seqsplit{execute\_code} \\
\midrule
Reasoning \& Coding Agent &
\seqsplit{execute\_code} \newline
\seqsplit{extract\_excel\_content} \newline
\seqsplit{extract\_document\_content} \\
\bottomrule
\end{tabular}
\caption{Worker agent implementation details for GAIA-MAPS, adapted
from the OWL framework~\cite{owl2025}.}
\label{tab:owl-tooling}
\end{table}

\subsection{MULTITAT Agent Setup Details}
\label{app:multitat-agent-setup}

For MULTITAT dataset, we use the lighter version based on the OWL implementation. We limit the worker agent to one as task was to extract the information from tables and text. Full MULTITAT agent details are---The planning, reasoning-coding agent, and final-answer agents, all use the \emph{model under evaluation}, which in a given run is one of Mistral-Large-3, or Qwen3-VL-235B-A22B. For Mistral-Large-3 and Qwen3-VL-235B-A22B the temperature is left at the provider default. Table~\ref{tab:multitat-config} summarizes
the agent-to-model assignment.

\begin{table}[H]
\centering
\footnotesize
\renewcommand{\arraystretch}{2.0}
\begin{tabular}{p{2.2cm}p{3.5cm}}
\hline
\textbf{Agent} & \textbf{Model} \\
\hline
Planning agent & model under evaluation \\
Reasoning and coding agent & model under evaluation \\
Final answer agent & model under evaluation \\
\hline
\end{tabular}
\caption{MULTITAT Agent configuration on MULTITAT dataset.}
\label{tab:multitat-config}
\end{table}

\section{GAIA-MAPS Quantitative Results}
\label{app:gaia-results}

\subsection{Run-to-Run Stability (GPT-5-mini)}
\label{app:gpt5-mini-gaia-results}

To assess robustness, we repeat the GPT-5-mini evaluation on GAIA-MAPS with an independent second run under identical settings. Table~\ref{tab:gaia-results-gpt5} reports per-language accuracy for Baseline and TART across both runs. \textbf{TART's improvement is consistent across runs, improving over Baseline in 20 of 22 language-run comparisons, with mean gains of +5.6 and +4.7 points in Run 1 and Run 2 respectively}, confirming the reported improvement is not an artifact of a single run. Due to compute constraints, this repeated-run analysis was conducted for GPT-5-mini only.

\begin{table}[H]
\centering
\footnotesize
\setlength{\tabcolsep}{4pt}
\renewcommand{\arraystretch}{1.8}
\begin{tabular}{@{}lrrrrr@{}}
\toprule
\textbf{Lang} & \textbf{Base} & \textbf{TART 1} & \textbf{Gain(R1)} & \textbf{TART 2} & \textbf{Gain(R2)} \\
\midrule
en & 38.78 & \textbf{43.63} & \textbf{\gain{4.85}}  & 40.60 & \textbf{\gain{1.82}} \\
ru & \textbf{32.12} & 31.51 & \textbf{\gain{-0.61}} & 32.12 & \textbf{\gain{0.00}} \\
de & 29.09 & 38.18 & \textbf{\gain{9.09}}  & \textbf{41.21} & \textbf{\gain{12.12}} \\
ar & 29.09 & \textbf{36.96} & \textbf{\gain{7.87}}  & 35.75 & \textbf{\gain{6.66}} \\
hi & 26.06 & 35.75 & \textbf{\gain{9.69}}  & \textbf{37.57} & \textbf{\gain{11.51}} \\
bn & 23.03 & \textbf{29.69} & \textbf{\gain{6.66}}  & 24.84 & \textbf{\gain{1.81}} \\
ky & 24.24 & 25.45 & \textbf{\gain{1.21}}  & \textbf{26.06} & \textbf{\gain{1.82}} \\
sw & 23.63 & 25.45 & \textbf{\gain{1.82}}  & \textbf{27.27} & \textbf{\gain{3.64}} \\
yo & 8.48  & \textbf{19.39} & \textbf{\gain{10.91}} & 13.93 & \textbf{\gain{5.45}} \\
ig & 18.18 & \textbf{26.06} & \textbf{\gain{7.88}}  & 22.42 & \textbf{\gain{4.24}} \\
ny & 20.60 & 23.03 & \textbf{\gain{2.43}}  & 23.03 & \textbf{\gain{2.43}} \\
\midrule
\textbf{Avg} & \textbf{24.85} & \textbf{30.46} & \textbf{\gain{5.61}} & \textbf{29.53} & \textbf{\gain{4.68}} \\
\bottomrule
\end{tabular}
\caption{Baseline vs. TART accuracy (EM) for GPT-5-mini across two independent runs on GAIA-MAPS, spanning 11 languages. Positive gains in \textcolor{green!60!black}{green}, negative in \textcolor{red}{red}.}
\label{tab:gaia-results-gpt5}
\end{table}

\subsection{Task Complexity and the Level-3 Plateau  }
\label{app:level-wise-analysis}

Figure~\ref{fig:experiments_y}(c) shows TART improves accuracy at Levels 1 and 2 but yields negligible gain at Level 3, averaged across all three models. To investigate this, Table~\ref{tab:level-wise-llm-jud} reports baseline failure categories by difficulty level from our LLM-judge analysis, aggregated across GPT-5-mini, Mistral-Large-3, and Qwen3-VL-235B-A22B on the seven-language subset used in Figure~\ref{fig:experiments} (a). Taxonomy-covered failure share is nearly identical between Levels 2 and 3---$55.2\%$ each, indicating TART has a comparable share of addressable planning-grounding failures to correct at both Levels 2 and 3.

\begin{table}[H]
\centering
\footnotesize
\renewcommand{\arraystretch}{1.8}
\begin{tabularx}{\linewidth}{@{}Xrrr@{}}
\toprule
\textbf{Level} & \textbf{Other} & \textbf{Total} & \textbf{Taxonomy-covered} \\
\midrule
Level 1 & 315 & 795 & $60.4\%$ \\
Level 2 & 652 & 1454 & $55.2\%$ \\
Level 3  & 212 & 473 & $55.2\%$ \\
\bottomrule
\end{tabularx}
\caption{Baseline failure counts by GAIA-MAPS difficulty level (LLM-judge), aggregated across all three models on the seven-language subset.}
\label{tab:level-wise-llm-jud}
\end{table}

Table~\ref{tab:level-wise-tool-steps} reports the average reference solution steps and required tools per task, from GAIA\cite{mialon2024gaia} ground-truth English task annotations — an intrinsic task property, independent of model, agent behavior, or language. Reference steps rise modestly from Level 1 to Level 2 (5.48 → 7.48) but nearly double from Level 2 to Level 3 (7.48 → 13.00); required tools follow the same trend (1.58 → 2.55 → 3.38).

\begin{table}[H]
\centering
\footnotesize
\renewcommand{\arraystretch}{1.8}
\begin{tabularx}{\linewidth}{@{}Xrrr@{}}
\toprule
\textbf{Level} & \textbf{Tasks} & \textbf{Avg. steps} & \textbf{Avg. tools} \\
\midrule
Level 1 & 53 & 5.48 & 1.58 \\
Level 2 & 86 & 7.48 & 2.55 \\
Level 3  & 26 & 13.00 & 3.38 \\
\bottomrule
\end{tabularx}
\caption{Average reference solution length and tool requirements by GAIA difficulty level \cite{mialon2024gaia}}
\label{tab:level-wise-tool-steps}
\end{table}

Since taxonomy-coverage is flat between Levels 2 and 3 while reference task length nearly doubles, the Level-3 plateau is better explained by intrinsic task complexity than by fewer correctable planning failures: even when TART resolves the initial planning-grounding failure and keeps the relevant constraints in view for the coordinator and worker agents, a longer downstream execution chain still gives more independent opportunities for retrieval, tool-use, or reasoning errors to occur. This finding indicates that Level 3's negligible improvement is driven by task complexity: as the number of required steps, tool calls, and retrieval operations expands significantly.

\subsection{Mistral and Qwen Quantitative Results}
\label{app:mis-qwen-gaia-results}

Table \ref{tab:results-mistral-qwen} reports per-language accuracy on GAIA-MAPS for Mistral-Large-3 and Qwen3-VL-235B-A22B across the seven shared languages. TART improves every language for Mistral-Large-3 (mean +5.97) and six of seven for Qwen3-VL (mean +3.55), with no degradation for either model, confirming that TART's benefit transfers across model families.

\begin{table}[H]
\centering
\footnotesize
\setlength{\tabcolsep}{3pt}
\renewcommand{\arraystretch}{3.0}
\begin{tabularx}{\linewidth}{@{}Xrrr rrr@{}}
\toprule
& \multicolumn{3}{c}{\textbf{Mistral-L3}} & \multicolumn{3}{c}{\textbf{Qwen3-VL}} \\
\cmidrule(lr){2-4}\cmidrule(lr){5-7}
\textbf{Lang.} & Base & TART & Gain & Base & TART & Gain \\
\midrule
English & 29.09 & 38.78 & \gain{9.69} & 29.69 & 33.33 & \gain{3.64} \\
German & 27.87 & 35.15 & \gain{7.28} & 29.69 & 30.30 & \gain{0.6} \\
Arabic  & 23.63 & 30.30 & \gain{6.67} & 21.81 & 25.45 & \gain{3.64} \\
Hindi   & 26.70 & 30.90 & \gain{4.24} & 26.06 & 30.90 & \gain{4.84} \\
Kyrgyz  & 16.36 & 23.03 & \gain{6.67} & 14.54 & 23.63 & \gain{9.09} \\
Yoruba  & 7.27 & 10.30 & \gain{3.03} & 9.69 & 9.69 & \gain{0.0} \\
Igbo  & 6.06 & 10.30 & \gain{4.24} & 7.87 & 10.90 & \gain{3.03} \\
\midrule
\textbf{Avg.} & 19.56 & 25.54 & \gain{5.97} & 19.91 & 23.46 & \gain{3.55} \\
\bottomrule
\end{tabularx}
\caption{Accuracy (\%) of baseline vs.\ TART for Mistral-Large-3 and
Qwen3-VL on GAIA-MAPS. Gain is TART $-$ baseline. TART improves accuracy on every language for Mistral-Large-3 (mean gain +5.97 points) and on six of seven languages for Qwen3-VL (mean gain +3.55 points), with no degradation observed for either model. Positive in
\textcolor{green!60!black}{green}, negative in \textcolor{red}{red}.}
\label{tab:results-mistral-qwen}
\end{table}

\section{MULTITAT: Mistral and Qwen Quantitative Results}
\label{app:multitat-results}

\begin{table}[H]
\centering
\footnotesize
\setlength{\tabcolsep}{3pt}
\renewcommand{\arraystretch}{1.8}
\begin{tabularx}{0.95\linewidth}{@{}Xrrr rrr@{}}
\toprule
& \multicolumn{3}{c}{\textbf{Mistral-L3}} & \multicolumn{3}{c}{\textbf{Qwen3-VL}} \\
\cmidrule(lr){2-4}\cmidrule(lr){5-7}
\textbf{Lang.} & Base & TART & Gain & Base & TART & Gain \\
\midrule
English & 25.60 & 33.20 & \gain{7.60} & 33.20 & 37.60 & \gain{4.40} \\
German & 20.40 & 31.60 & \gain{11.20} & 30.00 & 33.20 & \gain{3.20} \\
Russian & 20.40 & 34.40 & \gain{14.00} & 30.00 & 30.80 & \gain{0.80} \\
Japanese  & 21.20 & 29.20 & \gain{8.00} & 29.60 & 34.80 & \gain{5.20} \\
French  & 16.80 & 34.40 & \gain{17.60} & 32.00 & 37.20 & \gain{5.20} \\
Spanish  & 22.80 & 34.00 & \gain{11.20} & 32.80 & 34.80 & \gain{2.00} \\
Chinese  & 20.00 & 29.60 & \gain{9.60} & 29.20 & 30.80 & \gain{1.60} \\
Bengali  & 20.80 & 30.00 & \gain{9.20} & 31.20 & 35.20 & \gain{4.00} \\
Telugu  & 22.40 & 27.60 & \gain{5.20} & 29.60 & 32.40 & \gain{2.80} \\
Swahili  & 19.6 & 26.40 & \gain{6.80} & 29.60 & 30.80 & \gain{1.20} \\
\midrule
\textbf{Avg.} & 21.00 & 31.04 & \gain{10.04} & 30.72 & 33.76 & \gain{3.04} \\
\bottomrule
\end{tabularx}
\caption{Accuracy (\%) of baseline vs.\ TART for Mistral-Large-3 and
Qwen3-VL on MULTITAT. Gain is TART $-$ baseline; positive in
\textcolor{green!60!black}{green}, negative in \textcolor{red}{red}.}
\label{tab:mul-results-mistral-qwen}
\end{table}

\section{Quantifying The Failure: LLM as a Judge}
\label{label: LLM Judge}
\paragraph{Judge design:}We design a robust prompt for the judge LLM whose system prompt specifies the task definitions, the set of allowed failure categories, and an example of each failure type. The in-context examples are held fixed across all categorizations and are drawn from a diverse set of languages, so that the judge is calibrated against multilingual demonstrations rather than examples from any single language.

\paragraph{Judge input:}For each sample, we provide the judge with: the query in the target language (e.g., Igbo), the English reference query, the ground-truth answer, the agent's final answer, and the agent's plan, together with the per-category multilingual examples described above. Conditioned on this input, the judge assigns the sample to one of the failure categories in our taxonomy. 

\paragraph{Category assignment:}When more than one failure occurs within a single agent plan, we instruct the judge to mark a single \textit{primary} category, defined as the failure most responsible for the erroneous plan. This yields a mutually exclusive labeling scheme consistent with the convention used in our manual analysis, and avoids double-counting when reporting per-category and per-language distributions.

If a sample does not fall into any category of our taxonomy, the judge marks it as \textit{other}. This category captures two cases: a planning failure not covered by our taxonomy, or an execution error occurring after the planning step.

\paragraph{Model:}We use Anthropic's Claude Opus~4.8 ~\cite{claudeopus48} as the judge model for categorizing all samples, deployed via AWS Bedrock with \texttt{max\_tokens} set to 8192 and the temperature left at the model
provider's default. We apply the LLM judge only to baseline failed
samples.

\subsection{LLM Judge System Prompt}
\label{app:llm-judg-sys-pmt}

The full system prompt used to instruct the LLM Judge is shown below.

\begin{tcblisting}{
    colback=gray!5,
    colframe=black!60,
    listing only,
    breakable,
    boxrule=0.5pt,
    left=4pt, right=4pt, top=3pt, bottom=3pt,
    listing options={
        basicstyle=\ttfamily\footnotesize,
        breaklines=true,
        columns=fullflexible,
        breakatwhitespace=false,
        escapeinside={(*}{*)}
    }
}

You are an expert evaluator for multilingual agentic planning failures.

Your task is to categorize the planning failure. You will be given:
1. A target-language query.
2. An English reference query. This is only for semantic clarification; the target-language query is the main query.
3. The English plan produced by the agent from the target-language query.
4. The ground-truth answer.
5. The agent final answer.
6. A list of allowed failure categories with definitions.
7. Several labeled examples showing how categories should be assigned.

You must assign exactly one final category from the allowed categories. If the failure does not clearly fit any allowed category, assign "other".

Important evaluation rules:
- Focus primarily on planning failure, especially whether the agent plan preserves the intent and constraints of the target-language query.
- Use the English reference query only to clarify meaning when the target-language query is ambiguous.
- Do not assume a failure category only because the final answer is wrong. First check whether the plan itself shows the failure.
- If the plan is faithful and it satisfy all the entities & constraints mentioned in the target language query, categorize "other" unless one of the allowed categories explicitly covers that failure.
- If multiple categories seem possible, choose the category that is most causally responsible for the wrong answer.
- Do not over-penalize paraphrases. A plan can use different wording if it preserves the same entities, source, time constraints, operations, and answer format.
- Keep the explanation concise, paper-ready, and focused on what went wrong.
- Do not repeat the full question in the explanation.
- Output must be in English only.
- Output must be valid JSON only.
- Return only two fields: "final_category" and "explanation".

Allowed categories:

1. "entity_grounding_failure"
Use when the plan drops, mistranslates, replaces, or hallucinates key entities such as people, places, documents, species, objects, titles, organizations, or target items.
- Use this category for entity grounding failures, where the plan contains entities that are not present in the query.
- Use this category when plan contains hallucinated categories and it leads to  wrong operation chain & an irrelevant answer.
Example signals:
- The plan uses the wrong titles, species, person, place, object.
- The plan introduces an entity not mentioned or implied in the query. For eg: \icecream\ or \lawsuit\ not mentioned in a query anywhere.
- The Plan hallucinate the entities involved in a query.
- The plan fails to resolve an implicit entity correctly.
- The planner grounds the plan to an entity that is not mentioned in the query, leading to an incorrect execution chain.
- The plan drops and partly uses entities instead using all entities mentioned in the query.

2. "source_grounding_failure"
Use when the plan drops, corrupts, replaces, or misuses the required source.
Example signals:
- The query requires a specific file, spreadsheet, document, website, citation link, table, or image, but the plan uses another source.
- The plan mutates a file path or URL.
- The plan accesses the correct broad webpage but extracts evidence from the wrong part of the source, such as webpage text instead of an image, if source modality is central.
- The plan drops and partly uses a constraint source instead using full source with correct search criteria.

3. "temporal_grounding_failure"
Use when the plan drops, changes, ignores, or misapplies a required time constraint.
Example signals:
- The query says ``as of May 2023,'' ``before 2020,'' or ``in April 1977,'' but the plan does not preserve it.
- The plan applies the time constraint to the wrong entity, source, or operation.

4. "operation_grounding_failure"
Use when the plan follows the wrong execution or reasoning chain, even if some entities or constraints are preserved.
Example signals:
- The plan solves the task in the wrong order.
- The plan searches for the wrong type of evidence or performs the wrong operation, such as searching for a title inside a source when it should first identify a place and then perform a separate lookup.
- The plan changes the intended objective.

5. "answer_format_grounding_failure"
Use when the task intent is mostly solved but the final answer or plan fails to preserve the required answer format.
- use this category when final answer matches the ground truth answer but have addtional metadata.
Example signals:
- Use this category if final answer matches the ground truth answer but have extra text.
- The expected answer is only a number, name, city, ZIP code, string, or comma-separated list, but the agent adds extra explanation, provenance, year, director, citations, or metadata.
- The answer type is wrong, such as returning a paragraph when only a number is required.
- The content is correct but the output format makes it non-compliant.
- if final answer is correct but has extra text. for eg: approximately 41 of them would be false positives in their claims instead "41" as a number directly.

6. "other"
Use when:
- The failure does not clearly fit the above categories.
- The plan is faithful, perfect and preserved all the constraints.
- There is insufficient information to confidently assign one of the allowed categories.
- Target language query intent, entities, sources are different from english query.
- Final answer is not correct.

Category selection priority:
- If the plan uses the wrong entity/document/object, choose "entity_grounding_failure".
- If the plan corrupts or loses a required source, choose "source_grounding_failure".
- If the plan loses or misapplies a time condition, choose "temporal_grounding_failure".
- If the plan preserves correct entities but uses the wrong procedure or objective, choose "operation_grounding_failure".
- If the plan and reasoning are mostly correct but the final output violates the requested format, choose "answer_format_grounding_failure".
- If none clearly applies, choose "other".

Use the labeled examples below as guidance for category boundaries:

<LABELLED_EXAMPLES>
{{examples}}
</LABELLED_EXAMPLES>

Now evaluate the following case:

Target-language query:
{{target_language_query}}

English reference query:
{{english_reference_query}}

Agent plan produced from target-language query:
{{agent_plan}}

Ground-truth answer:
{{ground_truth_answer}}

Agent final answer:
{{agent_final_answer}}

Return only valid JSON in this exact format:

{
  "final_category": "one_of_the_allowed_categories",
  "explanation": "Concise explanation of why this category applies."
}

\end{tcblisting}

\subsection{LLM Judge Response Examples}
\label{label:llm-jud-resp}

\begin{qualexample}{\large\textbf{Example 1: Source Grounding Failure}}

\textbf{Target Language:} Arabic

\textbf{English Reference Query:} \textit{The attached spreadsheet contains a list of books I read in the year 2022. What is the title of the book that I read the slowest, using the rate of words per day? Here are the necessary table files: \textcolor{blue}{\textbf{da52d699-e8d2-4dc5-9191-a2199e0b6a9b}}.xlsx, for processing excel file, you can write python code and leverage excel toolkit to process the file step-by-step and get the information.}

\textbf{Ground Truth Answer:} \textit{Out of the Silent Planet}

\textbf{Agent Final Answer:} \textit{The title of the book read at the slowest rate (lowest words per day) in 2022 is: [title]}

\textbf{Agent Plan:} 1. Extract the content of the Excel file (da52d699-\textcolor{red}{\textbf{e812}}-4dc5-9191-a2199e0b6a9b.xlsx) to retrieve the list of books, their word counts, and the reading duration (start and end dates or total days). 

2. Write Python code to calculate the reading rate......(continue)

\textbf{LLM Judge Response:} The plan corrupts the required file path. The original identifier contains 'da52d699-e8d2', but the plan mutates it to 'da52d699-e812', changing 'e8d2' to 'e812'. This source-string mutation causes the tool to reference a non-existent Excel file, so the correct data cannot be accessed, and the agent returns an unfilled placeholder answer instead of 'Out of the Silent Planet'.
\end{qualexample}

\begin{qualexample}{\large\textbf{Example 2: Operation Grounding Failure}}

\textbf{Target Language:} Hindi

\textbf{English Reference Query:} \textit{As of May 2023, how many stops are between South Station and Windsor Gardens on MBTA’s Franklin-Foxboro line (not included)?}

\textbf{Ground Truth Answer:} \textit{10}

\textbf{Agent Final Answer:} \textit{12}

\textbf{Agent Plan:} 1. Search for the MBTA Franklin-Foxboro Line route map or official schedule as of May 2023, specifically focusing on the segment between South Station and Windsor Gardens, to identify all intermediate stops (excluding Franklin-Foxboro stations)

2. Extract and list all the stops located between South Station and Windsor Gardens on the Franklin-Foxboro Line as of May 2023, ensuring that Franklin-Foxboro branch stations are excluded as per the task requirement.

3. Count the number of stops identified in the previous step and verify the count using an alternative source or official MBTA documentation to ensure accuracy.

4. Transform the final count into the required answer format: \"The number of stops between South Station and Windsor Gardens on the MBTA Franklin-Foxboro Line (excluding Franklin-Foxboro branch stations) as of May 2023 is [X]. 

\textbf{LLM Judge Response:} The plan preserves the correct endpoints (South Station, Windsor Gardens), the source (MBTA Franklin-Foxboro Line), and the temporal constraint (May 2023). However, it misinterprets the '(not included)' constraint as excluding 'Franklin-Foxboro branch stations' rather than treating the endpoints themselves as non-inclusive when counting intermediate stops. This wrong operationalization of the exclusion rule leads to counting the wrong set of stops, producing 12 instead of the correct 10.
\end{qualexample}

\begin{qualexample}{\large\textbf{Example 3: Other Failure}}

\textbf{Target Language:} Hindi

\textbf{English Reference Query:} \textit{When you take the average of the standard population deviation of the red numbers and the standard sample deviation of the green numbers in this image using the statistics module in Python 3.11, what is the result rounded to the nearest three decimal points? Here are the necessary image files: df6561b2-7ee5-4540-baab-5095f742716a.png}

\textbf{Ground Truth Answer:} \textit{17.056}

\textbf{Agent Final Answer:} \textit{17.533}

\textbf{Agent Plan:} 1. Extract the red and green numbers from the image file located at "df6561b2-7ee5-4540-baab-5095f742716a.png".

2. Calculate the population standard deviation for the red numbers and the sample standard deviation for the green numbers using Python 3.11's statistics module.

3. Average the two computed standard deviations and round the result to three decimal places.

4. Verify the result by recalculating using an alternative method or tool to ensure accuracy.

5. Transform the final computed value into the required response format: \"The result is [value] when rounded to three decimal places.

\textbf{LLM Judge Response:} The plan faithfully preserves all entities (red numbers, green numbers), the correct source image, the correct operations (population std for red, sample std for green, average, round to three decimals), and the required answer format. The plan is correct and intent-preserving, yet the final numeric answer (17.533) does not match the ground truth (17.056), indicating an execution-level error in extracting or computing values rather than a planning grounding failure.
\end{qualexample}

\subsection{LLM Judge Quantitative Results}
\label{label:llm-judg-result-model}

We apply the LLM judge to the baseline failure samples across all three models on the GAIA-MAPS dataset. Language-wise and model-wise values are
reported in Table~\ref{tab:judge-results}.

\begin{table}[H]
\centering
\scriptsize
\setlength{\tabcolsep}{3pt}
\renewcommand{\arraystretch}{1.2}
\begin{tabularx}{\linewidth}{@{}l *{6}{Y} r@{}}
\toprule
\textbf{Lang.} & \textbf{C1} & \textbf{C2} & \textbf{C3} & \textbf{C4} & \textbf{C5} & \textbf{C6} & \textbf{Total} \\
\midrule
\multicolumn{8}{@{}l}{\textit{GPT-5-mini}} \\
English & 1 & 0 & 0 & 9 & 16 & 74 & 100 \\
German  & 2 & 1 & 1 & 7 & 31 & 74 & 116 \\
Arabic  & 8 & 2 & 0 & 18 & 21 & 67 & 116 \\
Hindi   & 6 & 0 & 1 & 21 & 27 & 65 & 120 \\
Kyrgyz  & 8 & 0 & 1 & 31 & 22 & 60 & 122 \\
Yoruba  & 22 & 1 & 8 & 57 & 25 & 36 & 149 \\
Igbo & 16 & 1 & 2 & 63 & 17 & 35 & 134 \\
\midrule
Total & 63 & 5 & 13 & 206 & 159 & 411 & 857 \\
\midrule
\multicolumn{8}{@{}l}{\textit{Mistral-Large-3}} \\
English & 2 & 2 & 2 & 15 & 21 & 73 & 115 \\
German & 1 & 5 & 1 & 15 & 16 & 80 & 118 \\
Arabic  & 10 & 10 & 0 & 34 & 15 & 55 & 124 \\
Hindi   & 7 & 4 & 1 & 23 & 16 & 68 & 119 \\
Kyrgyz  & 22 & 5 & 2 & 45 & 22 & 40 & 136 \\
Yoruba  & 29 & 0 & 10 & 98 & 7 & 8 & 152 \\
Igbo  & 30 & 3 & 1 & 105 & 10 & 4 & 153 \\
\midrule
Total & 101 & 29 & 17 & 335 & 107 & 328 & 917 \\
\midrule
\multicolumn{8}{@{}l}{\textit{Qwen3-VL}} \\
English & 1 & 1 & 0 & 9 & 11 & 93 & 115 \\
German & 2 & 1 & 1 & 9 & 12 & 90 & 115 \\
Arabic  & 7 & 2 & 0 & 22 & 8 & 88 & 127 \\
Hindi   & 2 & 1 & 2 & 16 & 9 & 90 & 120 \\
Kyrgyz  & 22 & 3 & 1 & 44 & 13 & 57 & 140 \\
Yoruba  & 28 & 2 & 14 & 83 & 8 & 13 & 148 \\
Igbo  & 32 & 5 & 4 & 93 & 7 & 9 & 150 \\
\midrule
Total & 94 & 15 & 22 & 276 & 68 & 440 & 915 \\
\bottomrule
\end{tabularx}
\caption{Failure category counts on baseline failure samples across languages and models (GAIA-MAPS). Categories: C1 -- Entity Grounding; C2 -- Source Grounding; C3 -- Temporal Grounding; C4 -- Operation Grounding; C5 -- Answer Format Grounding; C6 -- Other.}
\label{tab:judge-results}
\end{table}

\subsection{Human Annotation Study}
\label{label:annotation_study}
We prepare a comprehensive annotation guide that presents, for each sample, all inputs supplied to the judge LLM together with the judge's response, and asks annotators to verify the assigned category. Each of the seven annotators is assigned 40 samples spanning different languages, giving 280 annotated samples out of 917 total failed examples, or approximately 30\% of the data. While the LLM judge is applied to all models in our evaluation, this human study is conducted on Mistral-Large~3 across seven languages: English (en), German (de), Arabic (ar), Hindi (hi), Kyrgyz (ky), Igbo (ig), and Yoruba (yo). These languages are deliberately chosen to span the resource spectrum of Common Crawl statistics, covering high-, mid-, and low-resource languages.

Across 117 decisive human verifications, 5 different annotators confirmed the LLM judge in 104 cases (88.9\%; 95\% Wilson CI: 81.9--93.4), corresponding to substantial chance-corrected reliability ($\kappa=0.860$) - see table \ref{tab:human-judge-reliability}.

\begin{table}[t]
\centering
\scriptsize
\setlength{\tabcolsep}{3.5pt}
\begin{tabular}{lrr}
\toprule
Judge category & $n$ & Agree (\%) \\
\midrule
Entity          & 18 & 94.4 \\
Source          & 10 & 100.0 \\
Temporal        & 4  & 100.0 \\
Operation       & 30 & 83.3 \\
Answer format   & 21 & 85.7 \\
Other           & 34 & 88.2 \\
\midrule
\textbf{Overall} & \textbf{117} & \textbf{88.9} \\
\bottomrule
\end{tabular}
\caption{Human verification of the LLM-judge categories. Results use decisive
annotations only; six uncertain annotations are excluded. Overall agreement
was 88.9\% (95\% Wilson CI: 81.9--93.4), with Cohen's
$\kappa=0.860$ and macro-F1 $=0.906$.}
\label{tab:human-judge-reliability}
\end{table}

\section{Common Crawl Per Language Resource Availability}
\label{app:common-crawl-stats}

Table \ref{tab:cc-language-share} reports Common Crawl share for each evaluated language \citep{commoncrawl-languages}, our proxy for resource availability. Values span roughly five orders of magnitude, from English $+40.58\%$ to Nyanja $+0.0008\%$, confirming a genuine high-to-low resource spectrum.

\begin{table}[H]
\centering
\footnotesize
\begin{tabular}{@{}lrr@{}}
\toprule
\textbf{Language} & \textbf{ISO} & \textbf{CC share (\%)} \\
\midrule
English  & eng & 40.5782 \\
Russian  & rus & 6.8217 \\
German   & deu & 5.9862 \\
Arabic   & ara & 0.6548 \\
Hindi    & hin & 0.2223 \\
Bengali  & ben & 0.1096 \\
Kyrgyz   & kir & 0.0127 \\
Swahili  & swa & 0.0117 \\
Yoruba   & yor & 0.0016 \\
Igbo     & ibo & 0.0012 \\
Nyanja   & nya & 0.0008 \\
\bottomrule
\end{tabular}
\caption{Common Crawl language share (\%) for each of the eleven evaluated languages, from crawl CC-MAIN-2026-30 \citep{commoncrawl-languages}. Languages ordered high- to low-resource.}
\label{tab:cc-language-share}
\end{table}

\end{document}